\documentclass[12pt]{article}
\usepackage{epsfig}
\usepackage{color}
\usepackage{amssymb,amsmath}
\usepackage{graphicx}
\usepackage{here}
\usepackage{cite}
\usepackage{ulem}
\usepackage{bm}
\newcommand{\ba}{\begin{align}}
\newcommand{\ea}{\end{align}}

 \makeatletter
\def\alt{\mathrel{\mathpalette\gl@align<}}
\def\agt{\mathrel{\mathpalette\gl@align>}}
\def\gl@align#1#2{\lower.6ex\vbox{\baselineskip\z@skip\lineskip\z@
\ialign{$\m@th#1\hfil##\hfil$\crcr#2\crcr\sim\crcr}}} \makeatother

\begin{document}
\begin{flushright}
\end{flushright}
\vspace*{1.0cm}

\begin{center}
\baselineskip 20pt 
{\Large\bf 
Sneutrinos as Mixed Inflaton and Curvaton
}
\vspace{1cm}

{\large 
Naoyuki Haba$^a$, \ Tomo Takahashi$^b$ \ and \ Toshifumi Yamada$^a$
} \vspace{.5cm}

{\baselineskip 20pt \it
$^a$ Graduate School of Science and Engineering, Shimane University, Matsue 690-8504, Japan \\
$^b$ Department of Physics, Saga University, Saga 840-8502, Japan
}

\vspace{.5cm}

\vspace{1.5cm} {\bf Abstract} \end{center}

We investigate a scenario where the supersymmetric partners of two right-handed neutrinos (sneutrinos) work as mixed inflaton and curvaton,
 motivated by the fact that the curvaton contribution to scalar perturbations can reduce the tensor-to-scalar ratio $r$
 so that chaotic inflation models with a quadratic potential are made consistent with the experimental bound on $r$.
After confirming that the scenario evades the current bounds on $r$ and the scalar perturbation spectral index $n_s$, 
 we make a prediction on the local non-Gaussianity in bispectrum, $f_{\rm NL}$, and one in trispectrum, $\tau_{\rm NL}$. 
Remarkably, since the sneutrino decay widths are determined by the neutrino Dirac Yukawa coupling, 
 which can be estimated from the measured active neutrino mass differences in the seesaw model, 
 our scenario has a strong predictive power about local non-Gaussianities, 
 as they heavily depend on the inflaton and curvaton decay rates.
Using this fact, we can constrain the sneutrino mass from the experimental bounds on $n_s$, $r$ and $f_{\rm NL}$.

\thispagestyle{empty}

\newpage

\setcounter{footnote}{0}
\baselineskip 18pt
%

\section{Introduction}

Sneutrino inflation~\cite{sneutrinoinflation} is a scenario
 where the supersymmetric partner of a right-handed neutrino (sneutrino), which is gauge singlet and hence has no D-term potential,
 serves as an inflaton
 \footnote{
Recent works on models in which a sneutrino is considered as the inflaton are, e.g., Refs.~\cite{recent}.
}.
The scenario is attractive phenomenologically, as a quadratic potential in chaotic inflation~\cite{Linde:1983gd} corresponds to 
 a Majorana mass term for the right-handed neutrino and therefore the seesaw~\cite{seesaw} scale is related to primordial density fluctuations. 
Also, since the reheating process is controlled by the sneutrino decay width, which is determined by the neutrino Dirac Yukawa coupling,
 one may expect a connection between the active neutrino mass and the reheating temperature.
Unfortunately, however, chaotic inflation with a quadratic potential is excluded at 2$\sigma$ level 
 by BICEP2/Keck Array~\cite{bicep} and Planck~\cite{planck} experiments, 
 because the model predicts too large ($\sim0.13$) tensor-to-scalar ratio $r$.
Nevertheless, if another sneutrino works as a curvaton~\cite{Enqvist:2001zp,Lyth:2001nq,Moroi:2001ct}, contributing to the generation of scalar perturbations without affecting the inflationary dynamics,
 $r$ is reduced and sneutrino inflation can still be consistent with the experimental data.
The possibility that multiple sneutrinos are responsible for primordial fluctuations is attractive
because there must be at least two right-handed neutrinos to account for the neutrino oscillation data.

In this paper, we investigate a scenario in which two sneutrinos play the role of mixed inflaton and curvaton \cite{Langlois:2004nn,Lazarides:2004we,Moroi:2005kz,Moroi:2005np,Ichikawa:2008iq,Fonseca:2012cj,Enqvist:2013paa,Vennin:2015vfa,Fujita:2014iaa}, 
one sneutrino yielding a sufficient number of $e$-folds and both sneutrinos generating scalar perturbations.
It will be shown that our scenario yields values of $r$ and the scalar perturbation spectral index $n_s$ that 
 are consistent with the bound from BICEP2/Keck Array and Planck.
We further derive a prediction for the local-type non-Gaussianitiy in bispectrum, $f_{\rm NL}$, and one in trispectrum, $\tau_{\rm NL}$\footnote{
One can also consider $g_{\rm NL}$ in trispectrum. 
However, since the current bound on $g_{\rm NL}$ and the one expected in the near future are not stringent enough to 
constrain the model, we do not discuss it here.
}. 

Remarkably, the sneutrino decay widths, which are essential parameters in the generation of local non-Gaussianities in a mixed inflaton-curvaton scenario, are determined from the neutrino Dirac Yukawa coupling, and 
 this Yukawa coupling can be related to the measured active neutrino mass differences $\Delta m_{21}^2$ and $\vert\Delta m_{32}^2\vert$
 if the fermionic partners of the sneutrinos are involved in the seesaw mechanism.
Therefore, our scenario is predictive about local non-Gaussianities, once the initial vacuum expectation values (VEVs) and masses of sneutrinos are fixed.
(Assumptions on the lightest neutrino mass and 
 the complex-valued rotation matrix appearing in Casas-Ibarra parametrization~\cite{ci} are still necessary for making a prediction.)
Taking advantage of the predictive power, we constrain the masses of sneutrinos and their VEVs when the pivot scale exited the horizon, 
  from the bound on $n_s$ and $r$ from BICEP2/Keck Array and Planck~\cite{bicep,planck} as well as the one on local non-Gaussianities~\cite{Ade:2015ava},
 for both the normal and inverted hierarchies of active neutrino mass.
Prospects for improved constraints on the sneutrino masses and VEVs are also studied.

Previously, Ref.~\cite{ellis} has pursued the possibility of accounting for the Planck data with multiple sneutrinos, with two working as inflatons
 and one as a curvaton.
Our study is different in that we also focus on local non-Gaussianities, which have not been discussed quantitatively in Ref.~\cite{ellis}.
Also, we conduct a comparative study for the normal and inverted hierarchies of active neutrino mass, which is not found in that paper.
Ref.~\cite{senoguz} has considered the scenario of mixed inflaton-curvaton sneutrinos from a different motivation, that is, 
 to mitigate the gravitino overproduction problem in low-scale supersymmetry breaking models. 
Ref.~\cite{lin} has dealt with a similar but different scenario, where two sneutrinos act 
as curvatons and the inflaton is supplied from elsewhere.

This paper is organized as follows.
In Section~2, we define the model and express the sneutrino widths in terms of the active neutrino mass differences under a natural assumption.
In Section~3, we describe the methodology of our analysis using the $\delta N$ formalism, with which we compute scalar perturbations.
We then present the results of our numerical analysis and study their implications.
Section~4 summarizes the paper.
\\

\section{Model}
 
We consider an extension of the minimal supersymmetric standard model (MSSM) with three gauge-singlet chiral superfields
 with $R$-parity~$=-1$, which we denote by $N_1,N_2,N_3$.
To realize large field inflation in the presence of supergravity effects (i.e., to evade the $\eta$ problem), 
 we assume a K\"ahler potential of no-scale supergravity type~\cite{noscale,noscaleinflation}.
On the other hand, the superpotential is assumed to take the same form as in global supersymmetry case,
 in order not to lose connection between inflationary physics and the seesaw mechanism.
The K\"ahler potential in our model is given by
\begin{align} 
K &= -3 \, \log\left[ T+T^\dagger
-\frac{1}{3}\sum_{i=1}^3\tilde{N}_i^\dagger\tilde{N}_i\right]+\Phi_{\rm MSSM}^\dagger\Phi_{\rm MSSM},
\label{kaehler}
\end{align}
 in the unit where the reduced Planck mass is set as $M_P=1$.
Here, $T$ is a gauge-singlet modulus superfield, $\Phi_{MSSM}$ collectively denotes the chiral superfields of the MSSM, and
 the gauge fields are omitted.
The superpotential is given by
\begin{align} 
W &= W_{\rm MSSM}+\sum_{i=1}^3\frac{1}{2}\tilde{M}_i \, \tilde{N}_i^2+\sum_{i=1}^3\sum_{\alpha=e,\mu,\tau}\tilde{h}_{i\alpha} \, \tilde{N}_i H_u L_\alpha,
\label{sup}
\end{align}
 where $W_{\rm MSSM}$ is the superpotential of the MSSM, $\tilde{M}_1,\tilde{M}_2,\tilde{M}_3$ denote supersymmetric masses for right-handed neutrinos,
 and $\tilde{h}_{i\alpha}$ represents the neutrino Dirac Yukawa coupling, with $i=1,2,3$ labeling right-handed neutrinos and $\alpha=e,\mu,\tau$ being the flavor of lepton doublets.
Terms responsible for supersymmetry breaking and non-zero cosmological constant at present are possibly introduced with some modifications of Eq.~(\ref{sup}), which we do not discuss in this paper.
We comment that the K\"ahler potential Eq.~(\ref{kaehler}) possesses the symmetry
\begin{align} 
T&\to T+\sum_{i=1}^3\epsilon_i^*\tilde{N}_i, \ \ \ \ \ \tilde{N}_i \to \tilde{N}_i+\epsilon_i \ \ \ \ \ (\epsilon_1,\epsilon_2,\epsilon_3{\rm \ are \ small \ constants}),
\end{align}
 which is explicitly violated by the superpotential, but not by the gauge interactions since $\tilde{N}_i$'s are gauge singlets.
Thus, the K\"ahler potential Eq.~(\ref{kaehler}) is natural in the sense that quantum corrections do not change its structure
 in the limit of vanishing superpotential.

The relevant part of the action reads (summation over $i,j=1,2,3$ and $\alpha=e,\mu,\tau$ should be taken)
\begin{align} 
S &\supset \int{\rm d}^4x \sqrt{-g}
\left[ \ K_{i\bar{j}} \, \partial_\mu \tilde{N}_j^\dagger \partial^\mu \tilde{N}_i + K_{i\bar{j}} \, \bar{\psi}_{\tilde{N}_j} \bar{\sigma}^\mu \partial_\mu\psi_{\tilde{N}_i} 
+K_{i\bar{T}} \, \partial_\mu T^\dagger \partial^\mu \tilde{N}_i + K_{i\bar{T}} \, \bar{\psi}_{T} \bar{\sigma}^\mu \partial_\mu\psi_{\tilde{N}_i}+{\rm H.c.}\right.
\nonumber \\
&+K_{T\bar{T}} \, \partial_\mu T^\dagger \partial^\mu T + K_{T\bar{T}} \, \bar{\psi}_{T} \bar{\sigma}^\mu \partial_\mu\psi_{T}
\nonumber \\
&\left.-e^{K/2}\left\{ \, \frac{1}{2}\tilde{M}_i \, \psi_{\tilde{N}_i}\psi_{\tilde{N}_i} 
+\tilde{h}_{i\alpha} \, \psi_{\tilde{N}_i} H_u \psi_{L_\alpha}+\tilde{h}_{i\alpha} \, \tilde{N}_i \psi_{H_u} \psi_{L_\alpha}
+h_{i\alpha} \, \psi_{\tilde{N}_i} \psi_{H_u} L_\alpha
+{\rm H.c.} \, \right\}-V \ \right],
\label{actionpre}
\end{align}
where $V$ is the scalar potential given by
\begin{align} 
V&=e^K \left\{ \, (K^{-1})_{i\bar{j}}(W_i+K_iW)(W_j+K_jW)^\dagger+(K^{-1})_{i\bar{T}}(W_i+K_iW)(K_TW)^\dagger+{\rm H.c.}\right.
\nonumber \\
&\left. +(K^{-1})_{T\bar{T}}(K_TW)(K_TW)^\dagger-3\vert W\vert^2 \, \right\}.
\end{align}
Here, we have defined
\begin{align}
&K_i\equiv\frac{\partial K}{\partial \tilde{N}_i},\ \ K_T\equiv\frac{\partial K}{\partial T},
 \ \ K_{i\bar{j}}\equiv \frac{\partial^2 K}{\partial \tilde{N}_i \partial \tilde{N}_j^\dagger},
\ \ K_{i\bar{T}}\equiv\frac{\partial^2 K}{\partial \tilde{N}_i \partial T^\dagger}, \ \
K_{T\bar{T}}\equiv \frac{\partial^2 K}{\partial T \partial T^\dagger}, \ \ 
W_i\equiv\frac{\partial W}{\partial \tilde{N}_i}.
\nonumber
\end{align}
The derivatives of the K\"ahler potential are obtained from Eq.~(\ref{kaehler}) as
\begin{align} 
K_{i\bar{j}}&=e^{2K/3}\left\{\delta_{ij}\left(T+T^\dagger
-\frac{1}{3}\sum_{k=1}^3\tilde{N}_k^\dagger\tilde{N}_k\right)+\frac{1}{3}\tilde{N}_i^\dagger\tilde{N}_j\right\},
\ \ \
K_{i\bar{T}}=-e^{2K/3}\tilde{N}_i^\dagger,
\ \ \
K_{T\bar{T}}=3e^{2K/3}.
\nonumber
\end{align}
Due to the no-scale structure, the scalar potential reduces to
\begin{align} 
V=& \ e^K \left\{ \, (K^{-1})_{i\bar{j}}(W_i+K_iW)(W_j+K_jW)^\dagger+(K^{-1})_{i\bar{T}}(W_i+K_iW)(K_TW)^\dagger+{\rm H.c.}\right.
\nonumber \\
&\left. +(K^{-1})_{T\bar{T}}(K_TW)(K_TW)^\dagger-3\vert W\vert^2 \, \right\}
\nonumber \\
=& \ e^{2K/3}\sum_{i=1}^3\left\vert\frac{\partial W}{\partial \tilde{N}_i}\right\vert^2.
\label{scalarpot}
\end{align}

Now, we assume that the VEV of the K\"ahler potential Eq.~(\ref{kaehler}) is stabilized at some value as 
\footnote{
We do not discuss a mechanism for stabilizing the VEV of $K$, as it is beyond the scope of this phenomenological study.
For model building attempts, see, e.g., Refs.~\cite{goncharov1,goncharov2,sugra}
}
\begin{align} 
\langle K \rangle &=c \ .
\end{align}
Also, the following hierarchy of the VEVs is assumed to hold throughout the history of the Universe:
\begin{align} 
e^{-c/3} &\gg \vert\langle\tilde{N}_i\rangle\vert \ \ (i=1,2,3),
\label{vevhierarchy1} \\
e^{-c/3}+\frac{1}{3}\vert\langle\tilde{N}_i\rangle\vert^2 &\gg 
\frac{1}{3}\vert\langle\tilde{N}_i\rangle\vert\vert\langle\tilde{N}_j\rangle\vert \ \
{\rm for} \ \ \vert\langle\tilde{N}_i\rangle\vert > \vert\langle\tilde{N}_j\rangle\vert,
\label{vevhierarchy2}
\end{align}
 where Eq.~(\ref{vevhierarchy2}) is compatible with our sneutrino mixed inflaton-curvaton
 scenario provided two of the three canonically-normalized sneutrinos always have VEVs below the reduced Planck mass. 
Given Eqs.~(\ref{vevhierarchy1}) and (\ref{vevhierarchy2}), we may ignore kinetic mixings among the sneutrinos and the modulus, and thereby make an approximation with $K_{i\bar{j}}\simeq e^{c/3}\,\delta_{ij}$ and define the canonically-normalized superfields, $N_i$, as $N_i\equiv e^{c/6} \, \tilde{N}_i$.
In terms of the canonically-normalized fields, the phenomenologically relevant part of the action Eq.~(\ref{actionpre}) is rewritten as
\begin{align} 
S &\supset \int{\rm d}^4x \sqrt{-g}
\left[ \, \partial_\mu N_i^\dagger \partial^\mu N_i + \bar{\psi}_{N_i} \bar{\sigma}^\mu \partial_\mu\psi_{N_i} \right.
\nonumber \\
&-\left\{ \, \frac{1}{2}M_i \, \psi_{N_i}\psi_{N_i} 
+h_{i\alpha} \, \psi_{N_i} H_u \psi_{L_\alpha}+h_{i\alpha} \, N_i \psi_{H_u} \psi_{L_\alpha}
+h_{i\alpha} \, \psi_{N_i} \psi_{H_u} L_\alpha
+{\rm H.c.} \, \right\}
\nonumber \\
&- \left. \vert M_i N_i +h_{i\alpha}H_u L_\alpha\vert^2 \, \right],
\label{action}
\end{align}
 where we have redefined the masses as $M_i \equiv e^{c/6}\, \tilde{M}_i$, and the Yukawa coupling as $h_{i\alpha}\equiv e^{c/3}\, \tilde{h}_{i\alpha}$.

In the action Eq.~(\ref{action}), $N_1,N_2,N_3$ can be interpreted as the superfields of right-handed neutrinos,
 and $M_1,M_2,M_3$ as their supersymmetric masses, which are equal to right-handed neutrino Majorana masses.
We take $M_1,M_2,M_3$ to be real positive by phase redefinition and then relabel them such that $M_1<M_2<M_3$ holds.
On the basis of the seesaw mechanism, we relate Yukawa coupling $h_{i\alpha}$ to the active neutrino masses $m_1,m_2,m_3$ and Majorana masses $M_1,M_2,M_3$ as
\begin{align} 
h_{i\alpha} &= \frac{\sqrt{2}}{v}\begin{pmatrix} 
      \sqrt{M_1} & 0 & 0 \\
      0 & \sqrt{M_2} & 0 \\
      0 & 0 &\sqrt{M_3} \\
   \end{pmatrix}
   R_{3\times3}
   \begin{pmatrix} 
      \sqrt{m_1} & 0 & 0 \\
      0 & \sqrt{m_2} & 0 \\
      0 & 0 &\sqrt{m_3} \\
   \end{pmatrix}
   U_{PMNS},
   \label{yukawa}
\end{align}
 where $v\simeq246$~GeV, $U_{PMNS}$ is the neutrino flavor mixing matrix, and 
 $R_{3\times3}$ is an arbitrary $3\times3$ rotation matrix with complex angles in Casas-Ibarra parametrizaton~\cite{ci}.

Additionally, we identify the radial part of the scalar components of $N_1,N_2,N_3$, denoted by $\phi_i\equiv \sqrt{2}\vert N_i\vert$ $(i=1,2,3)$, with mixed inflaton and curvatons.
The mass of $\phi_i$ is simply given by $M_i$. 
To derive the decay rate of $\phi_i$, we note that
 $\phi_i$ particle is allowed to decay exclusively into the fermion and scalar components of lepton doublets $L_\alpha$'s and the up-type Higgs doublet $H_u$.
In general, however, some of the MSSM flat directions might acquire Planck-scale VEVs, 
 which endow $L_\alpha$'s and $H_u$ with large supersymmetric masses and thereby kinematically block the $\phi_i$ particle decay.
We will show that such blocking is not the case in our model with the K\"ahler potential Eq.~(\ref{kaehler}).
To see this, recall that since the MSSM superfields are canonical in Eq.~(\ref{kaehler}),
 any MSSM scalar field $\Phi_{\rm MSSM}$ gains the following mass term during inflation:
\begin{align} 
-S &\supset \int{\rm d}^4x \sqrt{-g} \, e^K\vert W\vert^2 \, \Phi_{\rm MSSM}^\dagger\Phi_{\rm MSSM}.
\end{align}
During inflation, one of $\phi_i$'s develops a transplanckian VEV of $\sim 10M_P$
 and hence we have $e^{K/2}\vert W\vert \sim (\sqrt{3}/2\sqrt{2})10H$ ($H$ denotes the Hubble rate), 
 which gives that the VEV of any MSSM scalar field attenuates to zero while inflation is taking place.
We have thus confirmed that the VEVs of MSSM flat directions are zero at the time of inflaton and curvaton decays.
We further assume that the MSSM $\mu$-term is negligible compared to $M_1,M_2,M_3$.
Then, the width of $\phi_i$ particle, denoted by $\Gamma_i$, is calculated by neglecting the mass of the final-state particles
 and is found to be (contraction over $i$ should not be taken)
\begin{align} 
\Gamma_i &= \frac{4M_i}{16\pi}(h h^\dagger)_{ii}
= \frac{4M_i^2}{8\pi v^2} \ \left[R_{3\times3} 
\begin{pmatrix} 
      m_1 & 0 & 0 \\
      0 & m_2 & 0 \\
      0 & 0 & m_3 \\
   \end{pmatrix}
R_{3\times3}^\dagger\right]_{ii\,{\rm component}} \ \ \ \ \ (i=1,2,3).
\label{width}
\end{align}

At this stage, $\Gamma_i$'s have ambiguity due to arbitrariness of the matrix $R_{3\times3}$.
In this paper, we introduce an \textit{ansatz} that the active neutrino mass and the neutrino Dirac Yukawa coupling possess
 the same hierarchical structure, and we thus take
\begin{align} 
R_{3\times3} &=     \begin{pmatrix} 
      1 & 0 & 0 \\
      0 & 1 & 0 \\
      0 & 0 & 1
   \end{pmatrix} \ {\rm for} \ {\rm normal \ hierarchy} \ (m_1<m_2<m_3),
\label{r-assumption1} \\
R_{3\times3} &=     \begin{pmatrix} 
      0 & 0 & 1 \\
      1 & 0 & 0 \\
      0 & 1 & 0
   \end{pmatrix} \ {\rm for} \ {\rm inverted \ hierarchy} \ (m_3<m_1<m_2),
\label{r-assumption2}
\end{align}
 which gives
\begin{align} 
\Gamma_3 &= \frac{4M_3^2}{8\pi v^2}m_3, \ \Gamma_2=\frac{4M_2^2}{8\pi v^2}m_2, \ \Gamma_1=\frac{4M_1^2}{8\pi v^2}m_1 \ \ \
{\rm for} \ {\rm normal \ hierarchy},
\label{width1}\\
\Gamma_3 &= \frac{4M_3^2}{8\pi v^2}m_2, \ \Gamma_2=\frac{4M_2^2}{8\pi v^2}m_1, \ \Gamma_1=\frac{4M_1^2}{8\pi v^2}m_3 \ \ \ 
{\rm for} \ {\rm inverted \ hierarchy}.
\label{width2}
\end{align}
This \textit{ansatz} is natural in that $\Gamma_i$'s are not altered significantly by deformations of $R_{3\times3}$,
 and even if Eqs.~(\ref{r-assumption1}) and (\ref{r-assumption2}) do not hold exactly, Eqs.~(\ref{width1}) and (\ref{width2}) remain good estimates in the majority of parameter space.
In terms of the active neutrino mass differences, Eqs.~(\ref{width1}) and (\ref{width2}) can be recast into the forms,
\begin{align} 
\Gamma_3 &= \frac{4M_3^2}{8\pi v^2}\sqrt{m_{\rm lightest}^2+\Delta m_{21}^2+\vert\Delta m_{32}^2\vert}, \ \Gamma_2=\frac{4M_2^2}{8\pi v^2}\sqrt{m_{\rm lightest}^2+\Delta m_{21}^2}, \ \Gamma_1=\frac{4M_1^2}{8\pi v^2}m_{\rm lightest}
\nonumber \\
&{\rm for} \ {\rm normal \ hierarchy},
\\
\Gamma_3&=\frac{4M_3^2}{8\pi v^2}\sqrt{m_{\rm lightest}^2+\vert\Delta m_{32}^2\vert}, \ \Gamma_2=\frac{4M_2^2}{8\pi v^2}\sqrt{m_{\rm lightest}^2+\vert\Delta m_{32}\vert^2-\Delta m_{21}^2}, \
\Gamma_1= \frac{4M_1^2}{8\pi v^2}m_{\rm lightest}
\nonumber \\
&{\rm for} \ {\rm inverted \ hierarchy},
\label{width3}
\end{align}
 where $m_{\rm lightest}$ is the unknown mass of the lightest active neutrino.
In this paper, we assume $m_{\rm lightest}^2\ll \Delta m_{21}$ and make an approximation that $m_{\rm lightest}=0$.
Also, we quote the following experimental values from Particle Data Group~\cite{pdg}:
 $\Delta m_{21}^2= 7.53\times10^{-5}~{\rm eV}^2$, and $\vert\Delta m_{32}^2\vert= 2.44\times10^{-3}~{\rm eV}^2$
 in the normal hierarchy case and $\vert\Delta m_{32}^2\vert= 2.51\times10^{-3}~{\rm eV}^2$ in the inverted hierarchy case.
\\

\section{Numerical analysis} 
 
\subsection{$\delta N$ formalism}

We employ $\delta N$ formalism~\cite{Starobinsky:1986fxa,Sasaki:1995aw,Sasaki:1998ug} to calculate the scalar perturbations.
We take the pivot scale (the scale for which experimental data are reported) as $k_*/a_0=0.05$~Mpc$^{-1}$, with $a_0$ being the scale factor at present.
The calculation proceeds as follows.
Hereafter, we restore the reduced Planck mass $M_P\simeq2.435\times 10^{18}$~GeV in the equations.

First, we consider an unperturbed FRW spacetime,
 where the metric is given by d$s^2 = {\rm d}t^2 - a^2(t){\rm d}x^2$ with $a(t)$ being the scale factor,
 the total energy density is denoted by $\rho(t)$, and the VEV of each sneutrino is denoted by $\phi_i(t)$ $(i=1,2,...,n_f; \ n_f$ is the number of sneutrinos).
The decay products of sneutrinos are relativistic during the period of interest
 and we denote their energy density by $\rho_\gamma(t)$.
When two sneutrinos (i.e., the case with $n_f=2$) are involved in the inflationary dynamics (the case for a general $n_f$ is straightforwardly obtained),
 the equation of motion for the sneutrinos and the Einstein equation read
\begin{align} 
&\ddot{\phi}_1(t) + \left( \, 3H(t)+\Gamma_1 \, \right)\dot{\phi}_1(t) + M_1^2 \, \phi_1(t) =0,
\label{scalareom1}  \\
&\ddot{\phi}_2(t) + \left( \, 3H(t)+\Gamma_2 \, \right)\dot{\phi}_2(t) + M_2^2 \, \phi_2(t) =0,
\label{scalareom2} \\
&\rho_\gamma(t) + 4H(t) \, \rho_\gamma(t) = \Gamma_1 \dot{\phi}_1^2(t)+\Gamma_2 \dot{\phi}_2^2(t),
\label{radiation} \\
&\rho(t) = \left(\frac{1}{2}\dot{\phi}_1^2(t)+\frac{1}{2}M_1^2 \, \phi_1^2(t)\right)+\left(\frac{1}{2}\dot{\phi}_2^2(t)+\frac{1}{2}M_2^2 \, \phi_2^2(t)\right)+\rho_\gamma(t),
\label{density} \\
&H^2(t) = \frac{\rho(t)}{3M_P^2},
\label{eineq}
\end{align}
where a dot represents the derivative with respect to time.
We compute the number of $e$-folds from $t=t_*$ to $t=t_{\rm final}$,
 where $t_*$ will later be identified with the time when the pivot scale exited the horizon:
\begin{align} 
\overline{N}(t_*,t_{\rm final})&=\int_{t_*}^{t_{\rm final}} {\rm d}t \ H(t).
\label{nefolds}
\end{align}

Now we consider perturbations.
After a coordinate transformation to make the metric ${\bm x}$-indepedent at $t=t_*$,
 we calculate the local number of $e$-folds at ${\bm x}$ and $t=t_{\rm final}$ in the perturbed spacetime, $N(t_*,t_{\rm final};{\bm x})$.
Then $N(t_*,t_{\rm final};{\bm x})-\overline{N}(t_*,t_{\rm final})$ is equal to the curvature perturbation at ${\bm x}$.
By choosing the uniform energy density slice, 
 we can derive the curvature perturbation on the uniform energy density slice at $t=t_{\rm final}$ and ${\bm x}$, $\zeta(t_{\rm final},\, {\bm x})$, as follows:
\begin{align}
\zeta (t_{\rm final}, \bm{x})  = N(t_*,t_{\rm final};{\bm x})-\overline{N}(t_*,t_{\rm final}).
\label{zeta}
\end{align}
If $t_{\rm final}$ is set at a time when all scalar fields have decayed,
 $\zeta$ on superhorizon scales is conserved after $t=t_{\rm final}$, giving the primordial density perturbations.

Scalar perturbations originate from quantum fluctuations of the scalar fields during inflation.
Fluctuations of the scalar fields are expressed as
\begin{align} 
\phi_i(t,{\bm x}) &= \phi_i(t)+\delta\phi_i(t,{\bm x}),
\nonumber \\
\delta \phi_i(t,{\bm x}) &= \int\frac{{\rm d}^3{\bm k}}{(2\pi)^3}e^{i{\bm k}\cdot{\bm x}} \ \delta\phi_{i,{\bm k}}(t) \ \ \ \ \ (i=1,2),
\label{pert}
\end{align}
 where $\phi_i(t)$ follows Eqs.~\eqref{scalareom1}--\eqref{eineq}.
We identify $t_*$ with the time when the mode for the pivot scale $k_*$ leaves the horizon, i.e. the time satisfying
$k_*/a(t_*) = H(t_*)$.
The two point function of the modes with ${\bm k}$ and ${\bm k'}$ at time $t=t_*$ is given by
\begin{align} 
\langle \delta\phi_{i,{\bm k}}(t_*) \, \delta\phi_{j,{\bm k'}}(t_*) \rangle
&= (2 \pi)^3  \, \delta_{ij} \, \delta^3({\bm k}+{\bm k'}) \,  \frac{2\pi^2}{ \vert {\bm k}\vert^3} \,\left( \frac{H(t_*)}{2\pi} \right)^2.
\label{quantum}
\end{align}
We evaluate $N(t_*,t_{\rm final}; \, {\bm x})$ by setting $t_{\rm final}$ at a time $t_{\rm decay}$ when all scalar fields have decayed.
Since we have $k_*/a(t) \ll H(t)$ several Hubble times after $t_*$,
 the spatial derivative of fields $\partial \phi_i(t,{\bm x})/\partial {\bm x}$ is negligible in the computation of $N(t_*,t_{\rm decay}; \, {\bm x})$.
Hence, $N(t_*,t_{\rm decay}; \, {\bm x})$ depends only on the VEVs and their time derivative at the initial time,
 $\phi_i(t_*,{\bm x})$ and $\dot{\phi}_i(t_*,{\bm x})$.
Moreover, the contributions from the time derivative $\delta\dot{\phi}_{i,{\bm k}}(t_*)$ to $N(t_*,t_{\rm decay};\, {\bm x})$
 are suppressed by $-\dot{H}(t_*)/H^2(t_*) \ll 1$ compared to those from the VEVs $\phi_{i,{\bm k}}(t_*)$.
Therefore, $N(t_*,t_{\rm decay}; \, {\bm x})$ is evaluated by Taylor expanding the unperturbed number of $e$-folds 
 $\overline{N}(t_*,t_{\rm decay})$ with respect to the unperturbed initial VEVs $\phi_i(t_*)$.
Defining $\phi_{i*}\equiv\phi_i(t_*)$ and making the dependence of $\overline{N}$ on $\phi_{i*}$ explicit, we find
\begin{align} 
&N(t_*,t_{\rm decay};\, {\bm x})
\nonumber \\
&=\sum_{i=1}^{2} \ \frac{\partial \overline{N}(t_*,t_{\rm decay};\, \phi_{1*},\phi_{2*})}{\partial\phi_{i*}}\delta\phi_i(t_*,{\bm x})
+ \frac{1}{2}\sum_{i,j=1}^{2} \ \frac{\partial^2 \overline{N}(t_*,t_{\rm decay};\, \phi_{1*},\phi_{2*})}{\partial\phi_{i*}\partial\phi_{j*}}\delta\phi_i(t_*,{\bm x})\delta\phi_j(t_*,{\bm x}) + .....
\label{deltan}
\end{align}
Likewise, the local energy density $\rho(t, {\bm x})$ is determined by the initial VEVs $\phi_{1*},\phi_{2*}$,
 and so is the uniform energy density slice.
The power spectrum of the curvature perturbation is defined by
\begin{align} 
\left\langle \zeta ({\bm k} ) \zeta ({\bm k'}) \right\rangle = (2\pi)^3 \delta ({\bm k}  + {\bm k'}) \frac{2\pi^2}{\vert {\bm k} \vert^3} P_\zeta (k). 
\label{P_zeta_def}
\end{align}
From Eqs.~(\ref{zeta}),~(\ref{quantum}) and (\ref{deltan}), $P_\zeta (k)$ at the pivot scale $k_*$ is calculated to be
\begin{align} 
P_\zeta(k_*) &=\sum_{i=1}^{2} \, \left(\frac{\partial \overline{N}(t_*,t_{\rm decay};\, \phi_{1*},\phi_{2*})}{\partial\phi_{i*}}\right)^2
\left(\frac{H(t_*)}{2\pi}\right)^2.
\label{pr}
\end{align}
The spectral index of scalar perturbations, $n_s$, is then derived as
\begin{align} 
n_s-1 &= \frac{{\rm d}\ln P_\zeta(k_*)}{{\rm d}\ln k_*}=\frac{1}{P_\zeta(k_*)}\frac{1}{H(t_*)}\frac{{\rm d}P_\zeta(k_*)}{{\rm d}t_*},
\label{ns}
\end{align}
 and the local-type nonlinearity parameter in bispectrum,
 $f_{\rm NL}$, and one in trispectrum, $\tau_{\rm  NL}$,
 are respectively given by (the derivatives are performed with $\rho(t_{\rm decay})$ fixed)
\begin{align} 
\frac{6}{5}f_{\rm NL} &= \frac{\sum_{i,j}\, (\partial^2 \overline{N}/\partial \phi_{i*}\partial \phi_{j*})
(\partial \overline{N}/\partial \phi_{i*})(\partial \overline{N}/\partial \phi_{j*})}
{\left\{\sum_l\, (\,\partial \overline{N}/\partial \phi_{l*})^2\right\}^2},
\label{fnl}\\
\tau_{\rm NL} &= \frac{\sum_{i,j}\, (\partial^2 \overline{N}/\partial \phi_{i*}\partial \phi_{j*})^2
(\partial \overline{N}/\partial \phi_{i*})(\partial \overline{N}/\partial \phi_{j*})}
{\left\{\sum_l\, (\,\partial \overline{N}/\partial \phi_{l*})^2\right\}^3}.
\label{taunl}
\end{align}
Reminding that the tensor power spectrum is simply given by $P_T(k_*)=(8/M_P^2)(H(t_*)/2\pi)^2$,
 the tensor-to-scalar ratio, $r$, is found to be
\begin{align} 
r&=\frac{8}{M_P^2}\frac{1}{\sum_i \, \left(\partial \overline{N}/\partial\phi_{i*}\right)^2}.
\end{align}

To confront the prediction with experimental data,
 we recall that $k_*$ satisfies $k_*/a_0=0.05$~Mpc$^{-1}$.
Assuming that no entropy production occurs after $t=t_{\rm decay}$,
 the value of $H(t_*)$ is determined from the following relation:
\begin{align} 
H(t_*) &= \frac{k_*}{a(t_*)}=\frac{k_*}{a_0} \ \frac{a_0}{a(t_{\rm decay})} \ \frac{a(t_{\rm decay})}{a(t_*)}
\nonumber \\
&= 0.05~{\rm Mpc}^{-1} \ \left(\frac{g_{S,{\rm eff}} \, T_{\rm decay}^3}{g_{S,{\rm eff},0} \, T_0^3}\right)^{1/3} \ 
\exp\left[\overline{N}(t_*,t_{\rm decay};\, \phi_{1*},\phi_{2*})\right],
\label{kst}
\end{align}
 where $g_{S,{\rm eff}}$ and $g_{S,{\rm eff},0}$ are the effective degrees of freedom for entropy density at $t=t_{\rm decay}$ and at present, respectively,
 $T_{\rm decay}$ is the temperature of the radiation at $t=t_{\rm decay}$ given by
\begin{align} 
\rho_\gamma(t_{\rm decay}) &= \frac{\pi^2}{30}g_{\rm eff} \, T_{\rm decay}^4,
\label{temperature}
\end{align}
 with $g_{\rm eff}$ being the effective degrees of freedom at $t=t_{\rm decay}$,
 and $T_0\simeq2.7255$~K is the CMB temperature at present.
We have $g_{S,{\rm eff},0}=43/11$ at present, and
 since the radiation at $t=t_{\rm decay}$ is a thermal bath composed of the MSSM particles, 
 we have $g_{S,{\rm eff}}=g_{\rm eff}=915/4$.
\\

\subsection{Procedure for numerical analysis}

We assume that the lightest sneutrino $\phi_1$ does not contribute to the dynamics of the inflationary expansion 
of the Universe or the generation of curvature perturbations, which corresponds to the case with $\phi_1 \ll \phi_2,  \phi_3$ and 
$M_1 \ll M_2, M_3$.
We relabel the other sneutrinos as $\phi\equiv\phi_3, \, \sigma\equiv\phi_2$, 
 and their masses and widths as $m_\phi\equiv M_3, \, m_\sigma\equiv M_2, \, \Gamma_\phi\equiv \Gamma_3, \, \Gamma_\sigma\equiv \Gamma_2$
(remind that $M_1<M_2<M_3$ by definition), and further write their VEVs when the pivot scale exited the horizon as $\phi_*\equiv\phi(t_*), \, \sigma_*\equiv\sigma(t_*)$.
We restrict ourselves to the case with $\phi_*>M_P$ and $\sigma_*<M_P$,
 which is in accord with the assumption Eq.~(\ref{vevhierarchy2}) 
 that has justified our approximation on the second derivative of the K\"ahler potential $K_{i\bar{j}}\simeq e^{c/3}\,\delta_{ij}$.
When $\phi_*>M_P$, $\sigma_*<M_P$ and $m_\phi > m_\sigma$,
 $\phi$ always drives an inflationary expansion while $\sigma$ does not
\footnote{
If $\sigma_* > M_P$, and if the K\"ahler potential were arranged appropriately, the secondary 
inflation driven by $\sigma$ could happen \cite{Langlois:2004nn,Moroi:2005kz,Ichikawa:2008iq,Dimopoulos:2011gb}.
However, we do not pursue such a possibility here.
}.

Our analysis proceeds as follows:
First, we take a set of trial values of the two sneutrino masses $m_\phi, m_\sigma$ and their VEVs $\phi_*,\sigma_*$.
The widths $\Gamma_\phi,\Gamma_\sigma$ are then uniquely determined by Eq.~(\ref{width3})
 (with $m_{\rm lightest}=0$ and Particle Data Group values for $\Delta m_{21}^2$ and $\Delta m_{32}^2$).
Now that the masses, widths and initial VEVs of the sneutrinos are specified, we
 numerically solve Eqs.~(\ref{scalareom1})--(\ref{eineq}) to obtain $\phi(t), \ \sigma(t)$.
We note that the widths computed field-theoretically in Eq.~(\ref{width3})
 are equal to the decay rates appearing in the equation of motion Eqs.~(\ref{scalareom1}) and (\ref{scalareom2}),
 as has been studied in Ref.~\cite{decayrate}.
We set $t_{\rm decay}$ at a time when $\sigma$ has decayed and the Universe is dominated by radiation
\footnote{
Since we assume $m_\phi> m_\sigma$ and the decay rates are given by Eqs.~\eqref{width1} and \eqref{width2}, $\phi$ always decays earlier than $\sigma$.
}.

We assume that the energy density of $\sigma$ is negligible during the slow-roll of $\phi$.
It follows that the evolution of the Universe before the end of slow-roll of $\phi$ is totally determined by the dynamics of $\phi$.
Thus, the Hubble rate when the scale $k_*$ exited the horizon is given by
\begin{align} 
H(t_*)^2 & \simeq \frac{m_\phi^2 \phi_*^2}{6M_P^2}.
\end{align}
Also, $\overline{N}(t_*,t_{\rm decay};\, \phi_*,\sigma_*)$ is easily computed by dividing it into two parts, 
 one from $t=t_*$ to a time before the end of slow-roll of $\phi$, and one after that time until $t=t_{\rm decay}$.
The former part is regulated by the dynamics of $\phi$ and is thus a function of $\phi_*$ only;
 the latter part involves the full dynamics of $\phi$, $\sigma$ and radiation from $\phi$ and $\sigma$ decays, but does not depend on $\phi_*$ 
 because the field configuration is independent of $\phi_*$ at a time before the end of slow-roll of $\phi$.
Therefore, we may write
 \begin{align} 
\overline{N}(t_*,t_{\rm decay};\, \phi_*,\sigma_*) &= \overline{N}(t_*,t_{\phi{\rm end}};\, \phi_*)
+\overline{N}(t_{\phi{\rm end}},t_{\rm decay};\, \sigma_*),
\label{nexpression}
 \end{align}
 with $t_{\phi{\rm end}}$ being a time before the end of slow-roll of $\phi$,
 which, in our analysis, we choose such that $\phi(t_{\phi{\rm end}})=\sqrt{2}M_P$ holds.
Since the dynamics of the inflationary expansion before $t=t_{\phi{\rm end}}$ is
 determined by the $\phi$ field with a quadratic potential,
 the first term on the right-hand side is expressed analytically as
\begin{align} 
\overline{N}(t_*,t_{\phi{\rm end}};\, \phi_*)&=\frac{\phi_*^2 - \phi_{\rm end}^2}{4M_P^2},
\ \ \ {\rm with}\ \ \phi_{\rm end}\equiv\phi(t_{\phi{\rm end}})=\sqrt{2}M_P.
\label{nefolds1}
\end{align}
On the other hand, $\overline{N}(t_{\phi{\rm end}},t_{\rm decay};\, \sigma_*)$ is calculated numerically through the formula,
\begin{align} 
\overline{N}(t_{\phi{\rm end}},t_{\rm decay};\, \sigma_*)&=\int_{t_{\phi{\rm end}}}^{t_{\rm decay}} {\rm d}t \ H(t),
\label{nefolds2}
\end{align}
 with the solution to Eqs.~(\ref{scalareom1})--(\ref{eineq}) inserted into $H(t)$.
We further perform a numerical calculation on the derivatives of $\overline{N}(t_{\phi{\rm end}},t_{\rm decay};\, \sigma_*)$
 with a fixed final total energy density,
\begin{align} 
N_\sigma &\equiv \left.\frac{\partial \overline{N}(t_{\phi{\rm end}},t_{\rm decay};\,\sigma_*)}{\partial \sigma_*}
\right\vert_{\rho(t_{\rm decay}) \, {\rm fixed}} \ ,
\nonumber \\
N_{\sigma\sigma}&\equiv
\left.\frac{\partial^2 \overline{N}(t_{\phi{\rm end}},t_{\rm decay};\,\sigma_*)}{\partial \sigma_*^2}
\right\vert_{\rho(t_{\rm decay}) \, {\rm fixed}} \ .
\end{align}
This is done by evaluating the number of $e$-folds for a fixed value of the initial VEV and slightly perturbed ones as follows:
\begin{align} 
\overline{N}(t_{\phi{\rm end}},t_{\rm decay};\, \sigma_*)&=\int_{t_{\phi{\rm end}}}^{t_{\rm decay}} {\rm d}t \ H(t;\sigma_*),
\nonumber \\
\overline{N}(t_{\phi{\rm end}},t'_{\rm decay};\, \sigma_*+\Delta\sigma)&=\int_{t_{\phi{\rm end}}}^{t'_{\rm decay}} {\rm d}t \ H(t;\sigma_*+\Delta\sigma) \ \ \ {\rm with} \ \rho(t'_{\rm decay};\,\sigma_*+\Delta\sigma)=\rho(t_{\rm decay};\sigma_*),
\nonumber \\
\overline{N}(t_{\phi{\rm end}},t''_{\rm decay};\, \sigma_*+2\Delta\sigma)&=\int_{t_{\phi{\rm end}}}^{t''_{\rm decay}} {\rm d}t \ H(t;\sigma_*+2\Delta\sigma) \ \ \ {\rm with} \ \rho(t''_{\rm decay};\,\sigma_*+2\Delta\sigma)=\rho(t_{\rm decay};\sigma_*),
\end{align} 
  where $t'_{\rm decay}$, $t''_{\rm decay}$ are the times at which the energy density for perturbed VEVs equals that for the unperturbed VEV.
Here, $\rho(t_{\rm decay};\sigma_*)$ is numerically obtained from the solution to Eqs.~(\ref{scalareom1})--(\ref{eineq}).
Then we take the differences to find
\begin{align} 
N_\sigma &= \frac{\overline{N}(t_{\phi{\rm end}},t'_{\rm decay};\, \sigma_*+\Delta\sigma)-\overline{N}(t_{\phi{\rm end}},t_{\rm decay};\, \sigma_*)}{\Delta\sigma},
\nonumber \\
N_{\sigma\sigma} &= \frac{\overline{N}(t_{\phi{\rm end}},t''_{\rm decay};\, \sigma_*+2\Delta\sigma)
-2\overline{N}(t_{\phi{\rm end}},t'_{\rm decay};\, \sigma_*+\Delta\sigma)
+\overline{N}(t_{\phi{\rm end}},t_{\rm decay};\, \sigma_*)}{\Delta\sigma^2}.
\end{align}

Although we numerically compute $N_\sigma$ and $N_{\sigma\sigma}$ as described above, 
 it is insightful to show the analytical formulas which are derived under 
 the so-called sudden decay approximation \cite{Sasaki:2006kq}:
\begin{align} 
N_\sigma &= \frac{2 r_{\rm dec}}{3 \sigma_\ast}, \label{Nsig_formula1} \\
N_{\sigma\sigma} &= \frac{2 r_{\rm dec} }{9 \sigma^2_\ast} \left( 3 - 4 r_{\rm dec} -2 r_{\rm dec}^2 \right),
\label{Nsig_formula}
\end{align}
where  $r_{\rm dec}$ is given by
\begin{align} 
r_{\rm dec} = \left. \frac{3\rho_\sigma }{ 3 \rho_\sigma + 4  \rho_\gamma }  \right|_{\rm dec} 
\simeq \left( \frac{\sigma_\ast}{M_{\rm pl}} \right)^2 \sqrt{\frac{m_\sigma}{\Gamma_\sigma}},
\label{r_dec}
\end{align}
with $\rho_\sigma$ being the energy density of $\sigma$.
$r_{\rm dec}$ roughly represents the ratio of the energy density of $\sigma$ to the total one at the time of its decay.

With $N_\sigma$, the scalar perturbation power spectrum is recast into the following form that allows easy computation:
 \begin{align} 
P_\zeta(k_*) &= \left(\frac{\phi_*^2}{4M_P^4}+N_\sigma^2\right) \left( \frac{H_\ast}{2\pi} \right)^2
\simeq \left(\frac{\phi_*^2}{4M_P^4}+N_\sigma^2\right)\frac{m_\phi^2 \phi_*^2}{24\pi^2M_P^2}.
\label{pr2}
\end{align}
Since $m_\sigma^2\sigma_*^2\ll m_\phi^2\phi_*^2$ and $m_\sigma<m_\phi$,
 the dependence of $\sigma_*$ on $t_*$ is negligible compared to that of $\phi_*$, i.e.
  $({\rm d}\sigma_*/{\rm d}t_*)/({\rm d}\phi_*/{\rm d}t_*)=m_\sigma^2\sigma_*/(m_\phi^2\phi_*)\ll1$.
Therefore, Eq.~(\ref{ns}) is simplified as
\begin{align} 
n_s-1 &= -\frac{1}{\frac{\phi_*^2}{4M_P^4}+N_\sigma^2}
\left(\frac{1}{M_P^2}+\frac{1}{H(t_*)}\frac{{\rm d}}{{\rm d}t_*}N_\sigma^2\right)-\frac{4M_P^2}{\phi_*^2}
\nonumber \\
&\simeq -\frac{1}{\frac{\phi_*^2}{4M_P^2}+N_\sigma^2M_P^2}-\frac{4M_P^2}{\phi_*^2}.
\label{nsth}
\end{align}
$r$, $f_{\rm NL}$ and $\tau_{\rm NL}$ are straightforwardly obtained as
\begin{align} 
r &= \frac{8}{\frac{\phi_*^2}{4M_P^2}+N_\sigma^2M_P^2},
\label{rth}\\
\frac{6}{5}f_{\rm NL} &= \frac{\frac{1}{2M_P^2}\frac{\phi_*^2}{4M_P^4}+N_{\sigma\sigma}N_\sigma^2}
{\left( \, \frac{\phi_*^2}{4M_P^4}+N_\sigma^2 \, \right)^2},
\label{fnlth}\\
\tau_{\rm NL} &= \frac{\frac{1}{4M_P^4}\frac{\phi_*^2}{4M_P^4}+N_{\sigma\sigma}^2N_\sigma^2}
{\left( \, \frac{\phi_*^2}{4M_P^4}+N_\sigma^2 \, \right)^3}.
\end{align}

We settle the value of $m_\phi$ by requiring that the scalar perturbation amplitude calculated in Eq.~(\ref{pr2}) be consistent with the Planck observation~\cite{planck} at 1$\sigma$ level,
\begin{align}
P_\zeta(k_*)&=\left(\frac{\phi_*^2}{4M_P^4}+N_\sigma^2\right)\frac{m_\phi^2 \phi_*^2}{24\pi^2M_P^2}
=e^{3.094\pm0.034}\times10^{-10}.
\label{amplitude}
\end{align}
Practically, this is achieved by repeating the numerical calculation of $\overline{N}(t_{\phi{\rm end}},t_{\rm decay};\sigma_*)$, $N_\sigma$
 and the final total energy density $\rho(t_{\rm decay})$
 and solving Eq.~(\ref{amplitude}) for $m_\phi$ iteratively,
 until $m_\phi$ yields the scalar perturbation amplitude within the $1\sigma$ range.
In every numerical calculation, we determine $\phi_*$ from the number of $e$-folds through the following formula:
\begin{align} 
H(t_*) &= 0.05~{\rm Mpc}^{-1} \ \left(\frac{g_{\rm eff}^{1/4} \, \{(30/\pi^2)\rho(t_{\rm decay})\}^{3/4}}{g_{S,{\rm eff},0} \, T_0^3}\right)^{1/3} \ 
\exp\left[\frac{\phi_*^2 - \phi_{\rm end}^2}{4M_P^2}+\overline{N}(t_{\phi{\rm end}},t_{\rm decay};\, \sigma_*)\right].
\label{kst2}
\end{align}
In this way, we obtain the spectral index $n_s$, tensor-to-scalar ratio $r$, and local non-Gaussianities $f_{\rm NL}$ and $\tau_{\rm NL}$
 at the time when the mode $k_*$ exited the horizon, for a scalar perturbation amplitude consistent with the observation.
Since $m_\phi$ and $\phi_*$ are now fixed, the remaining free parameters are $m_\sigma$ and $\sigma_*$. 

Finally, the results are confronted with the bound on $(n_s,r)$ reported by BICEP2/Keck Array and Planck in Refs.~\cite{bicep,planck}, and the bound on $f_{\rm NL}$ reported by Planck in Ref.~\cite{Ade:2015ava},
 to constrain $(m_\sigma,\sigma_*)$.
\\

\subsection{Results of numerical analysis}
  
\begin{figure}[H]
  \begin{center}
    \includegraphics[width=80mm]{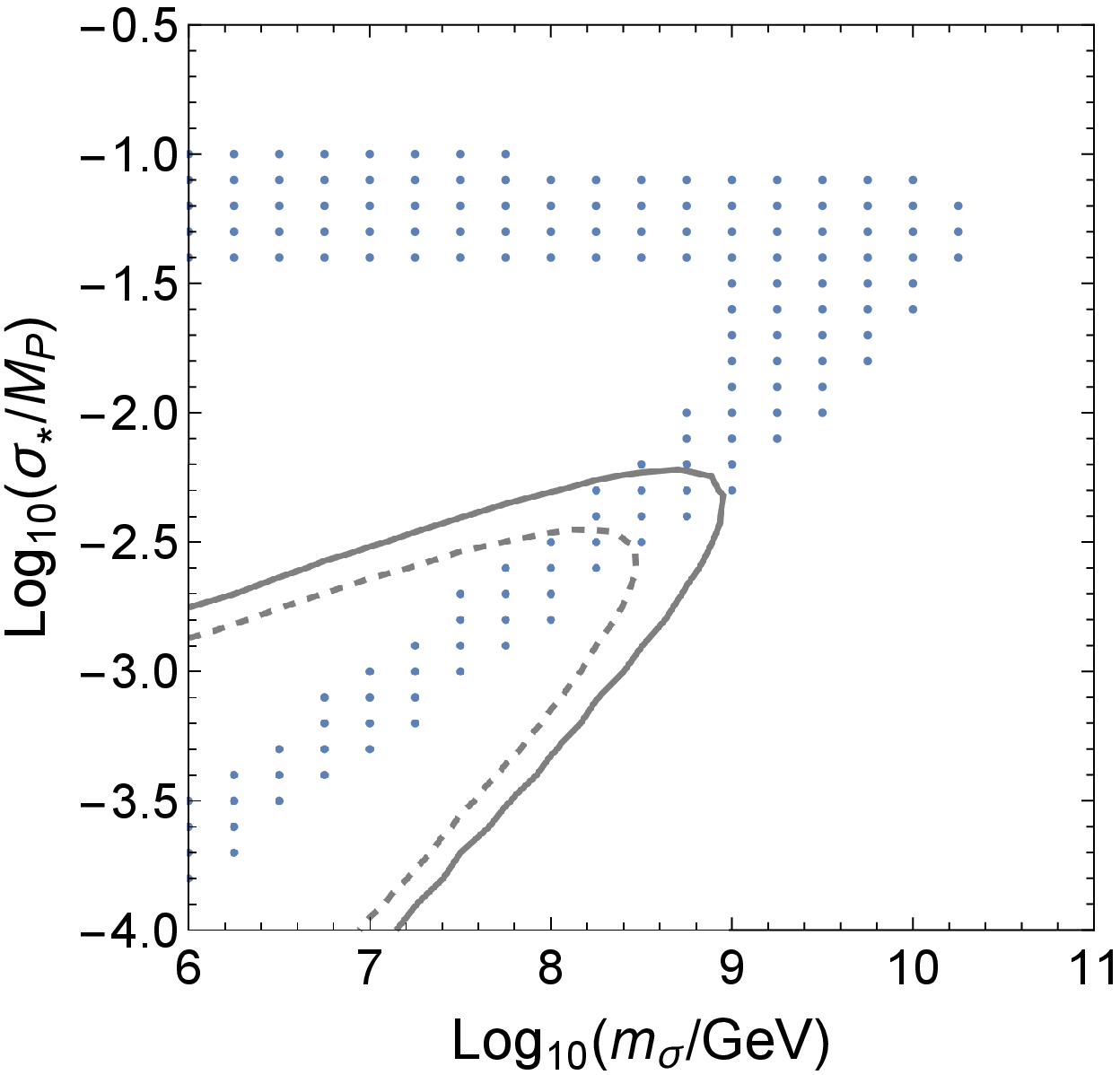} 
    \includegraphics[width=80mm]{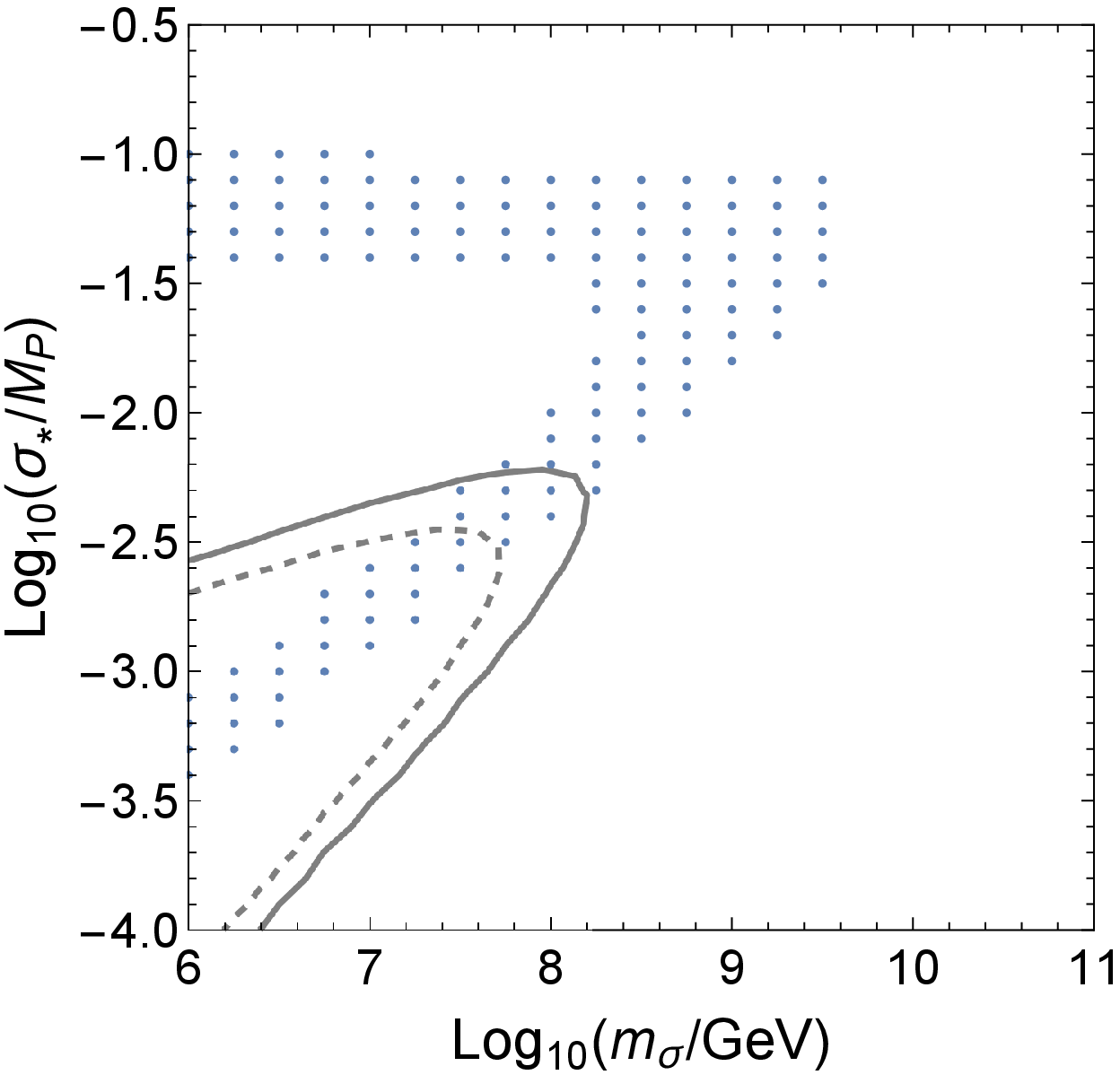}
    \caption{
    Contour plots on the plane of the second heaviest sneutrino mass $m_\sigma$, and its VEV when the pivot scale exited the horizon $\sigma_*$.
    The left panel is for the normal hierarchy of the active neutrino mass, and the right panel is for the inverted hierarchy.
    The area with dots represents the parameter region that is consistent, at 2$\sigma$ level, with the constraint on $(n_s,r)$ reported by  BICEP2/Keck Array+Planck~\cite{bicep,planck}.
    The solid and dashed contours respectively represent $f_{\rm NL} = 5.8$ and $10.8$, which correspond to 
1$\sigma$ and 2$\sigma$ upper bounds on the local non-Gaussianity $f_{\rm NL}$ reported by Planck~\cite{Ade:2015ava}, which reads $f_{\rm NL}=0.8\pm5.0$.
         }
    \label{contours}
  \end{center}
\end{figure}

In Figure~\ref{contours}, we present contour plots
 on the plane of the second heaviest sneutrino mass $m_\sigma$, and its VEV when the pivot scale exited the horizon $\sigma_*$.
Here, the area with dots represents the parameter region 
that is consistent, at 2$\sigma$ level, with the constraint on $(n_s,r)$ reported by BICEP2/Keck Array and Planck~\cite{bicep,planck}
 (no parameter region is consistent at 1$\sigma$).
The solid and dashed contours respectively indicate 1$\sigma$ and 2$\sigma$ upper bounds 
 on the local non-Gaussianity $f_{\rm NL}$ reported by Planck~\cite{Ade:2015ava}, which reads $f_{\rm NL}=0.8\pm5.0$.
From Figure~\ref{contours}, the following observations are made:
\begin{itemize}
\item 
By and large, there are two regions on $(m_\sigma,\,\sigma_*)$ plane that satisfy the constraint on $(n_s,r)$, which we label as (A) and (B);
(A) is where $m_\sigma\lesssim10^{10}$~GeV and $0.03M_P\lesssim\sigma_*\lesssim0.1M_P$,
 and (B) is where $\log m_\sigma$ and $\log \sigma_*$ are linearly correlated and again $m_\sigma\lesssim10^{10}$~GeV.
In both regions, $\sigma$'s contribution to scalar perturbations is of the same order as $\phi$'s contribution,
 i.e. $N_\sigma M_P \sim \phi_*/2M_P$ holds in Eq.~(\ref{pr2}),
 in which case $r$ is mildly reduced below the BICEP2/Keck Array+Planck bound and at the same time $n_s$ is not displaced much 
 from the central value of the BICEP2/Keck Array+Planck data.
For example, when $N_\sigma^2 M_P^2\simeq 1.2~\phi_*^2/4M_P^2$ holds, $r$ and $n_s$ are estimated as follows:
From the numerical calculation, we have found that the term $\phi_*^2/4M_P^2$ in Eq.~(\ref{nefolds1}) accounts for most of the number of $e$-folds and thus satisfies $\phi_*^2/4M_P^2\sim60$. Therefore,
\begin{align} 
r&=\frac{8}{\frac{\phi_*^2}{4M_P^2}+N_\sigma^2 M_P^2}\simeq0.06, \ \ \ \ \ 
n_s=1-\frac{1}{\frac{\phi_*^2}{4M_P^2}+N_\sigma^2M_P^2}-\frac{4M_P^2}{\phi_*^2}\simeq0.975,
\end{align}
 which are consistent with the BICEP2/Keck Array+Planck results at 2$\sigma$ level.

\item
In region (B), however,
 large $f_{\rm NL}$ is predicted and thus most of the parameter space is excluded by the Planck data.
The origin of large local non-Gaussianity is understood as follows.
As the initial VEV $\sigma_*$ decreases, the fraction of $\sigma$'s contribution to the total energy density when $\sigma$ decays
 becomes suppressed, and so does $r_{\rm dec}$ in Eq.~(\ref{r_dec}).
Since $N_\sigma M_P \sim \phi_*/2M_P$ holds in region (B) and thus $N_\sigma$ lies in a fixed range,
 we find from Eqs.~(\ref{Nsig_formula1}) and (\ref{Nsig_formula}) that $6f_{\rm NL}/5\sim3/(8r_{\rm dec})$ holds for $r_{\rm dec}\ll1$.
Therefore, large $f_{\rm NL}$ is predicted in the part of region (B) with smaller $\sigma_*$.

\item
The allowed region shifts towards left (smaller $m_\sigma$) in the inverted hierarchy case compared to the normal hierarchy case.
For instance, when $\sigma_*=0.05~M_P$, the allowed range of the second heaviest sneutrino mass is $m_\sigma <  2.0\times 10^{10}~{\rm GeV}$ in the normal hierarchy case, and $m_\sigma<5.6\times 10^9~{\rm GeV}$ in the inverted hierarchy case.
When $\sigma_*=0.01~M_P$, the allowed range is $5.0\times 10^8~{\rm GeV}<m_\sigma<4.0\times 10^9~{\rm GeV}$ in the normal hierarchy case, and $8.9\times 10^7~{\rm GeV}<m_\sigma<7.1\times 10^8~{\rm GeV}$ in the inverted hierarchy case.
This shift is because the width of $\sigma$ is an essential factor for determining the scalar perturbation power spectrum,
 and it is proportional to $m_\sigma^2 \sqrt{\Delta m_{21}^2}$ in the normal hierarchy case, 
 and to $m_\sigma^2 \sqrt{\vert\Delta m_{32}^2\vert-\Delta m_{21}^2}$ in the inverted hierarchy case.
Since $\vert\Delta m_{32}^2\vert \gg \Delta m_{21}^2 $, the same width is obtained with a smaller $m_\sigma$ in the inverted hierarchy case.
 
\end{itemize}

\begin{figure}[H]
  \begin{center}
    \includegraphics[width=80mm]{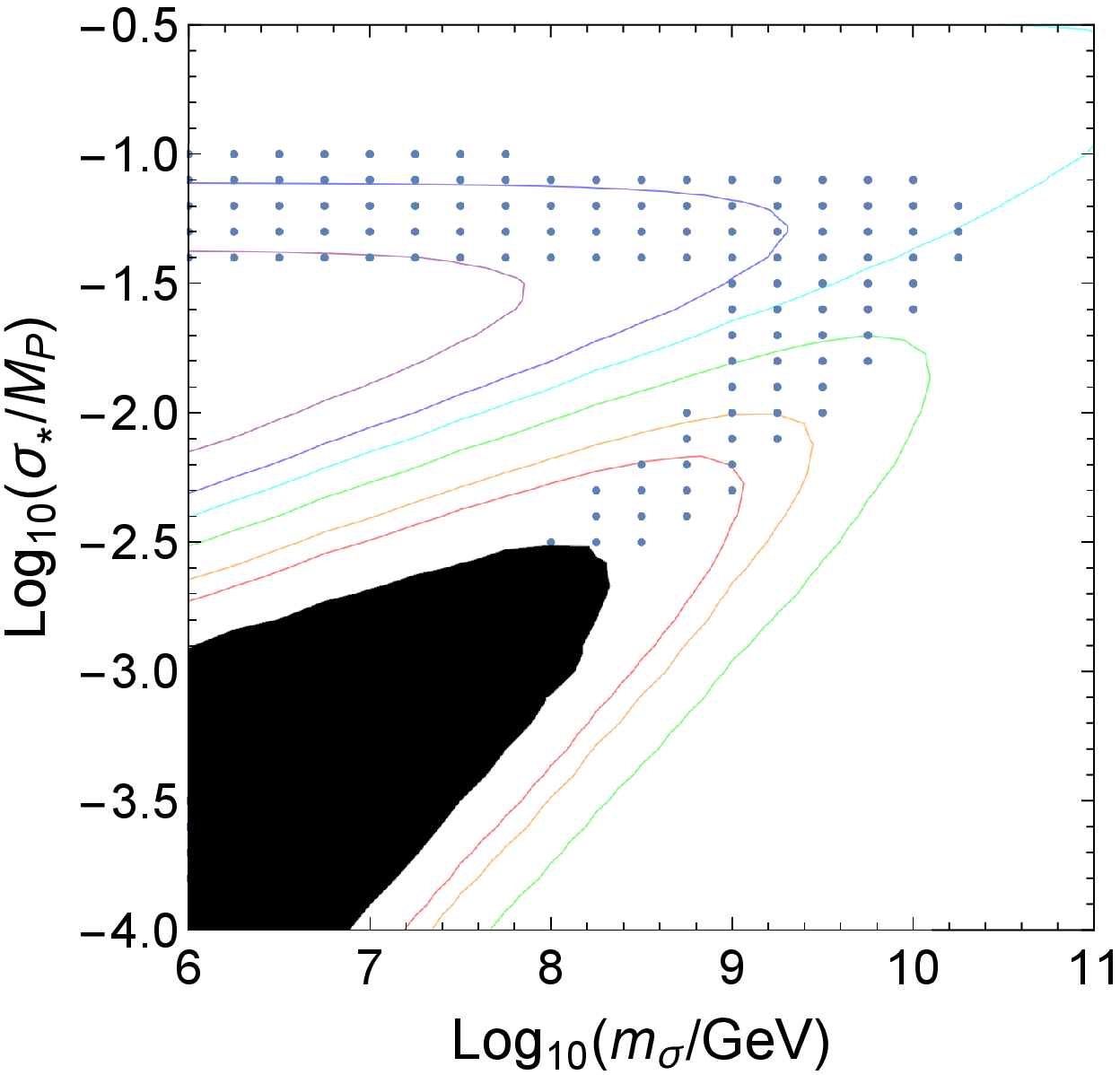} 
    \includegraphics[width=80mm]{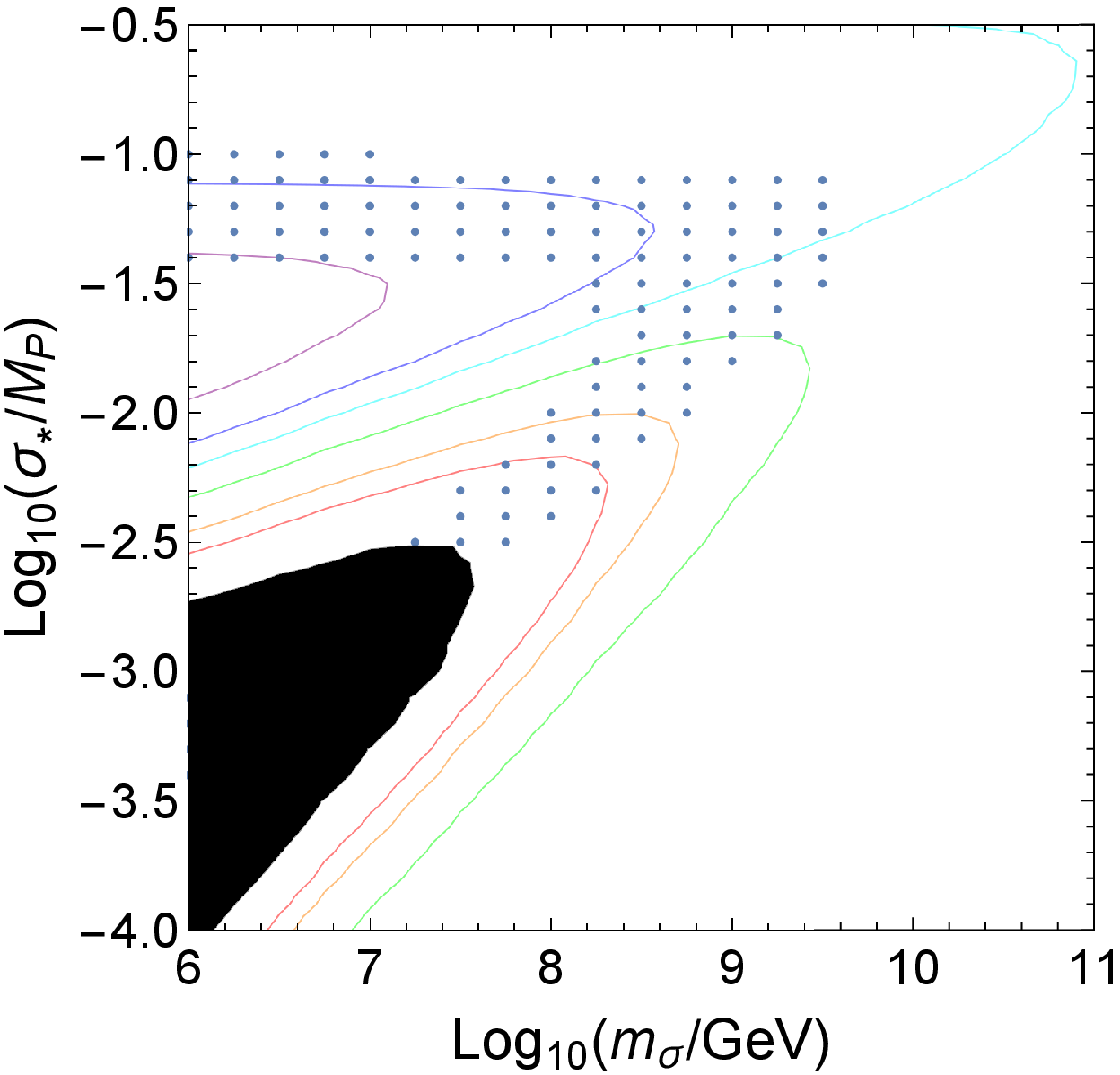}
    \caption{
    Contours of local non-Gaussianity $f_{\rm NL}$ on the plane of the second heaviest sneutrino mass $m_\sigma$, and its VEV when the pivot scale exited the horizon $\sigma_*$.
    $f_{\rm NL}=-1,\,-0.5,\,0,\,1,\,3,\,5$ on the purple, blue, light blue, green, orange and red contours, respectively.
    The left panel is for the normal hierarchy of the active neutrino mass, and the right panel is for the inverted hierarchy.
    The area with dots represents the parameter region that is consistent, at 2$\sigma$ level, with the constraint on $(n_s,r)$ reported by BICEP2/Keck Array+Planck~\cite{bicep,planck},
    and the black-filled area is the parameter region excluded by the Planck bound on $f_{\rm NL}$~\cite{Ade:2015ava} at 2$\sigma$ level.     
         }
    \label{fnlprediction}
  \end{center}
\end{figure}

\begin{figure}[H]
  \begin{center}
    \includegraphics[width=80mm]{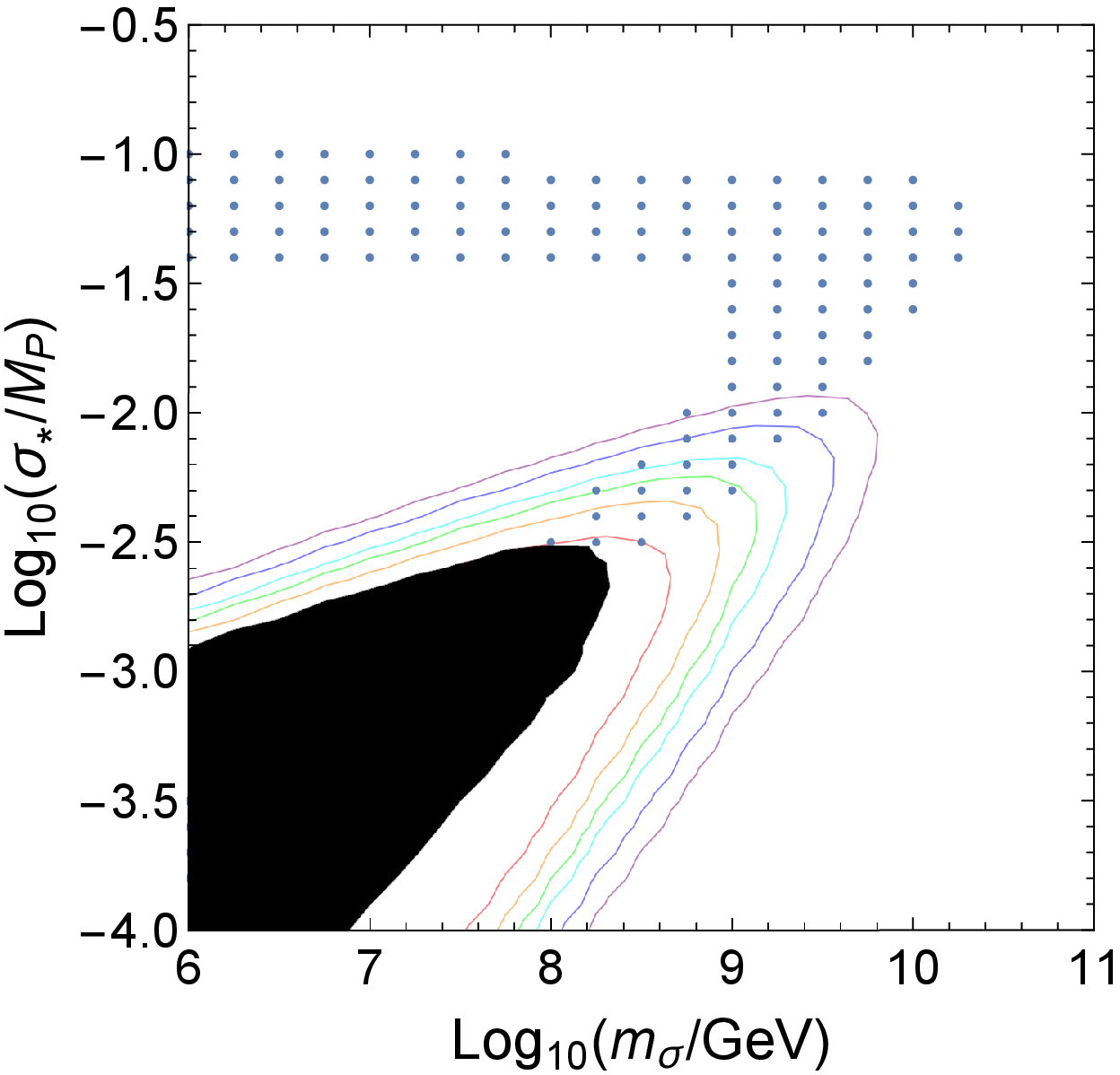} 
    \includegraphics[width=80mm]{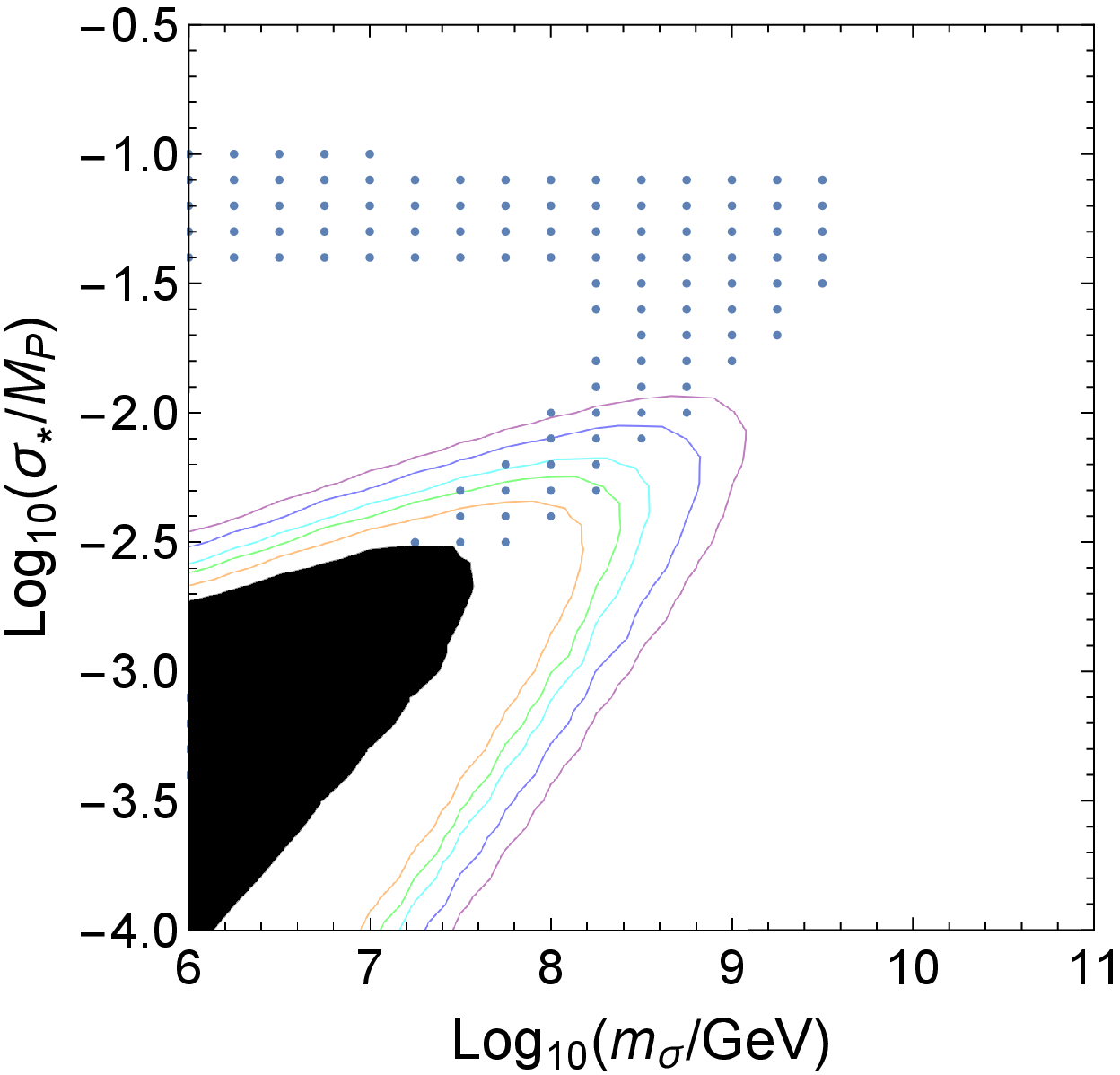}
    \caption{
    Contours of local non-Gaussianity $\tau_{\rm NL}$ on the plane of the second heaviest sneutrino mass $m_\sigma$, and its VEV when the pivot scale exited the horizon $\sigma_*$.
    $\tau_{\rm NL}=10,~20,~40,~60,~100,~200$ on the purple, blue, light blue, green, orange and red contours, respectively.
    The left panel is for the normal hierarchy of the active neutrino mass, and the right panel is for the inverted hierarchy.
    The area with dots represents the parameter region that is consistent, at 2$\sigma$ level, with the constraint on $(n_s,r)$ reported by BICEP2/Keck Array+Planck~\cite{bicep,planck},
    and the black-filled area is the parameter region excluded by the Planck bound on $f_{\rm NL}$~\cite{Ade:2015ava} at 2$\sigma$ level.     
         }
    \label{taunlprediction}
  \end{center}
\end{figure}

We make a prediction for local non-Gaussianities $f_{\rm NL}$ and $\tau_{\rm NL}$ in the allowed region, which are presented in Figure~\ref{fnlprediction} and Figure~\ref{taunlprediction}, respectively.
Since it is expected that future observations such as SKA and Euclid can probe  $f_{\rm NL}$ 
with an accuracy of $\Delta f_{\rm NL}\lesssim 1$~\cite{Yamauchi:2014ioa,Fonseca:2015laa,Yamauchi:2015mja}, we may be able
to distinguish region (A) from region (B), because $-0.5\gtrsim f_{\rm NL}\gtrsim -1$ is predicted in region (A), while $f_{\rm NL}\gtrsim 0$ in region (B).
Also, if region (B) is the case, $(m_\sigma,\sigma_*)$ can be restricted to a small area through the measurement of $f_{\rm NL}$.
In addition to $f_{\rm NL}$, $\tau_{\rm NL}$ may also provide useful information about $m_\sigma$ in region (B).
The combination of future observations of SKA and Euclide may probe $\tau_{\rm NL}$ with $\tau_{\rm NL} / \sigma (\tau_{\rm NL}) \ge 1$  where $\sigma (\tau_{\rm NL})$ is  1$\sigma$ uncertainty for $\tau_{\rm NL}$, for a wide range of $f_{\rm NL}$ and $\tau_{\rm NL}$.
For example, non-zero values of $\tau_{\rm NL}$ can be confirmed for the fiducial values of $f_{\rm NL} \simeq 1.5$ and $\tau_{\rm NL} \simeq 17$~\cite{Yamauchi:2015mja}.


\begin{figure}[H]
  \begin{center}
    \includegraphics[width=80mm]{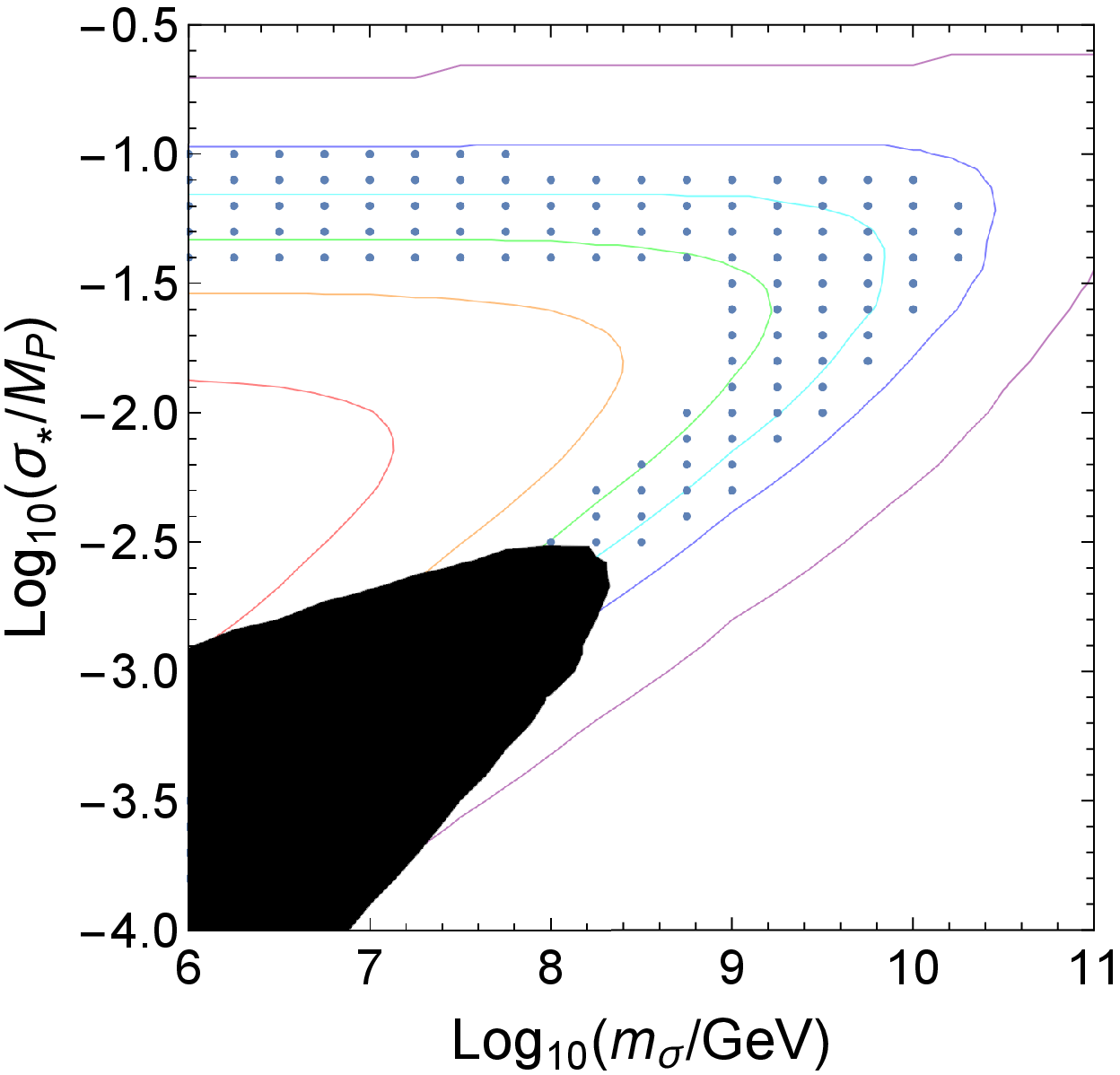} 
    \includegraphics[width=80mm]{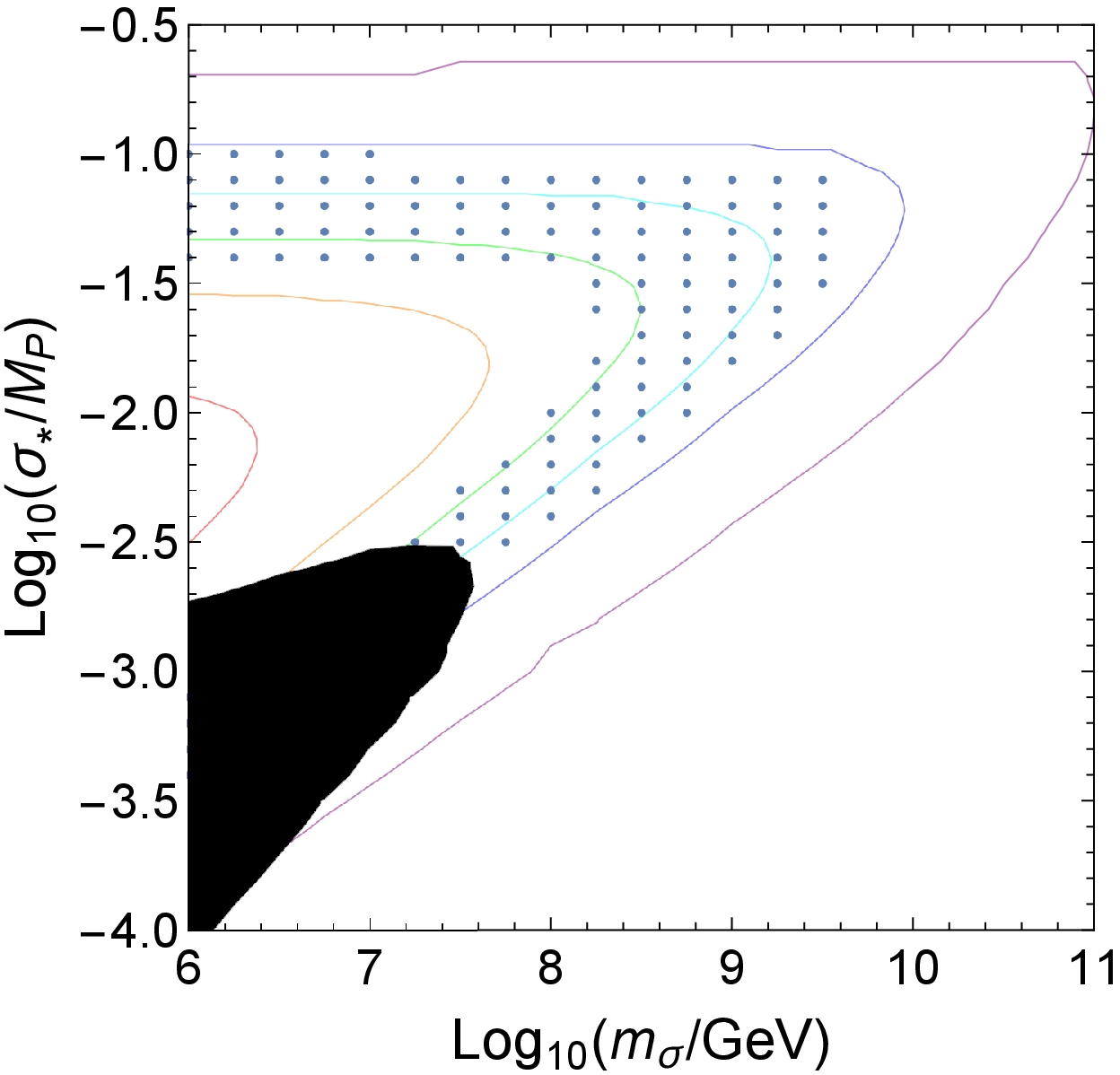}
    \caption{
    Contours of the largest right-handed neutrino Majorana mass $m_\phi(=M_3)$ on the plane of the second heaviest sneutrino mass $m_\sigma$, and its VEV when the pivot scale exited the horizon $\sigma_*$.
    $m_\phi=2.5\times10^{12},\,5\times10^{12},\,7.5\times10^{12},\,10^{13},\,1.25\times10^{13},\,1.5\times10^{13}$~GeV on the red, orange, green, light blue, blue and purple contours, respectively.
    The left panel is for the normal hierarchy of the active neutrino mass, and the right panel is for the inverted hierarchy.
    The area with dots represents the parameter region that is consistent, at 2$\sigma$ level, with the constraint on $(n_s,r)$ reported by BICEP2/Keck Array+Planck~\cite{bicep,planck},
    and the black-filled area is the parameter region excluded by the Planck bound on $f_{\rm NL}$~\cite{Ade:2015ava} at 2$\sigma$ level.     
         }
    \label{mphiprediction}
  \end{center}
\end{figure}

Finally, we present in Figure~\ref{mphiprediction} a prediction for the largest right-handed neutrino Majorana mass $m_\phi(=M_3)$,
 which satisfies the constraints from the number of $e$-folds Eq.~(\ref{kst2}) and the measurement of scalar perturbation amplitude Eq.~(\ref{amplitude}).
We find that in all parameter regions compatible with the BICEP2/Keck Array and Planck data,
 the value of $m_\phi$ is restricted to a narrow range of $0.75\times10^{13}$~GeV$\lesssim m_\phi\lesssim1.25\times10^{13}$~GeV.
This is because $N_\sigma M_P \sim \phi_*/2M_P$ must hold to fulfill the bound on $(n_s,r)$,
 and when this is the case, Eqs.~(\ref{kst2}) and (\ref{amplitude}) almost uniquely fix the values of $\phi_*$ and $m_\phi$,
 since the number of $e$-folds is mostly controlled by $\phi_*^2$ part of Eq.~(\ref{kst2}).

We discuss the reheating temperature and its phenomenological implications.
The temperature of radiation just after $\sigma$ decay, which we regard as the reheating temperature, is estimated to be
\footnote{
Remind that the temperature corresponding to $\rho(t_{\rm decay})$ derived from the solution to Eqs.~\eqref{scalareom1}--\eqref{eineq} does not have a physical meaning,
 because it depends on an arbitrary choice of $t_{\rm decay}$ in a numerical calculation.
This choice, of course, does not alter the final results, since $\rho(t_{\rm decay})^{1/4}e^{\overline{N}}$ in Eq.~(\ref{kst2})
 is invariant under a change of $t_{\rm decay}$.
}
\begin{align} 
T_{\rm reheating}\simeq \left(\frac{30}{\pi^2 g_{\rm eff}}3M_P^2\Gamma_\sigma^2\right)^{1/4}
&=3.3~m_\sigma \ \ \ {\rm for \ the \ normal \ hierarchy},
\\
&=7.9~m_\sigma \ \ \ {\rm for \ the \ inverted \ hierarchy}.
\end{align}
In the parameter region considered in Figure~\ref{contours}, the reheating temperature is above $10^6$~GeV
 and therefore the sphaleron process can take place to convert lepton number asymmetry to baryon number asymmetry.
We note that the correct amount of lepton number asymmetry for leptogenesis can always be produced in sneutrino decays
 by tuning the neutrino Majorana phases~\cite{sneutrinoinflation}.
\\

\section{Quantum corrections to the K\"ahler potential}

We check that quantum corrections to the K\"ahler potential do not spoil the slow-roll of sneutrino inflaton and curvaton
 (in this section, we use the unit with $M_P=1$).
Remind that the no-scale K\"ahler potential Eq.~(\ref{kaehler}) receives quantum corrections through the sneutrino superpotential couplings in Eq.~(\ref{sup})
 that break the no-scale structure.
The logarithmically divergent part of the one-loop corrections from the superpotential has been calculated in Ref.~\cite{gaillard} for generic supergravity theory.
In our case with the tree-level K\"ahler potential given by Eq.~(\ref{kaehler}), it reads
\footnote{
For a more general form of the K\"ahler potential
$K= -n\log[ T+T^\dagger-\sum_{i=1}^3\tilde{N}_i^\dagger\tilde{N}_i/n]+\Phi_{\rm MSSM}^\dagger\Phi_{\rm MSSM}$,
 it becomes
\begin{align} 
\Delta K&=\frac{\log\Lambda^2}{32\pi^2}\left(e^{(1-2/n)K}\sum_{i,j=1}^3\vert W_{ij} \vert^2+\left(2n-6+\frac{2}{n}\right)e^{(1-1/n)K}\sum_{i=1}^3\vert W_i\vert^2
+(n^2-4n-3)e^K\vert W\vert^2\right.
\nonumber\\
&+2e^{(1-1/n)K}\sum_{a={\rm MSSM}}\sum_{i=1}^3\vert\Phi_a^\dagger W_i+W_{ai}\vert^2
+(2n-2)e^K\sum_{a={\rm MSSM}}\vert W_a+\Phi_a^\dagger W\vert^2
\nonumber\\
&\left.+e^K\sum_{a,b={\rm MSSM}}\vert \Phi^\dagger_a W_b+\Phi^\dagger_b W_a + \Phi^\dagger_a\Phi^\dagger_b W+W_{ab}\vert^2\right).
\end{align}
}
\begin{align} 
\Delta K&=\frac{\log\Lambda^2}{32\pi^2}\left(e^{K/3}\sum_{i,j=1}^3\vert W_{ij} \vert^2+\frac{2}{3}e^{2K/3}\sum_{i=1}^3\vert W_i\vert^2
-6e^K\vert W\vert^2\right.
\nonumber\\
&+2e^{2K/3}\sum_{a={\rm MSSM}}\sum_{i=1}^3\vert\Phi_a^\dagger W_i+W_{ai}\vert^2
+4e^K\sum_{a={\rm MSSM}}\vert W_a+\Phi_a^\dagger W\vert^2
\nonumber\\
&\left.+e^K\sum_{a,b={\rm MSSM}}\vert \Phi^\dagger_a W_b+\Phi^\dagger_b W_a + \Phi^\dagger_a\Phi^\dagger_b W+W_{ab}\vert^2\right).
\label{correctionstokaehler}
\end{align}
 where $\Lambda$ is the cutoff scale, and $\Phi_a,\Phi_b$ denote scalar fields in MSSM.
Here, summation over $i,j$ is taken for the three right-handed neutrinos, and summation over $a,b$
 is for all MSSM scalar fields. 
$W_i$, $W_{ij}$, $W_a$, $W_{ab}$, $W_{ai}$ are defined as
 $W_i\equiv \partial W/\partial\tilde{N}_i$, $W_{ij}\equiv \partial^2 W/\partial\tilde{N}_i\partial\tilde{N}_j$,
 $W_a\equiv \partial W/\partial\Phi_a$, $W_{ab}\equiv \partial^2 W/\partial\Phi_a\partial\Phi_b$,
 $W_{ai}\equiv \partial^2 W/\partial\Phi_a\partial\tilde{N}_i$, respectively.

After the VEV of the K\"ahler potential is stabilized as $\langle K\rangle=c$, the following quadratic and quartic terms of sneutrinos
 are obtained from Eq.~(\ref{correctionstokaehler}):
\begin{align} 
&\Delta K\supset
\nonumber \\
&\frac{\log\Lambda^2}{32\pi^2}\left(
e^{c/3}\sum_{i,j=1}^3(hh^\dagger)_{ij} \,\tilde{N}^\dagger_j\tilde{N}_i
+\frac{2}{3}e^{c/3} \sum_{i=1}^3 M_i^\dagger M_i \, \tilde{N}^\dagger_i\tilde{N}_i
-6e^{2c/3} \sum_{i,j=1}^3 \frac{1}{4}M_i^\dagger M_j \, \tilde{N}^\dagger_i\tilde{N}^\dagger_i\tilde{N}_j\tilde{N}_j \right).
\label{correctionstonn}
\end{align}
 where $h_{i\alpha}=e^{c/3}\tilde{h}_{i\alpha}$ and $M_i=e^{c/6}\tilde{M}_i$ while the normalization of the sneutrino is unchanged.
Numerically, we have $(hh^\dagger)_{33} \gtrsim 2\vert\Delta m^2_{23}\vert M_3/v^2 > 10^{-3}$ and $M_i < 10^{-5}$ (see Figure~\ref{mphiprediction}) in the viable parameter region of our scenario, 
 and we also find $\langle \tilde{N}_i \rangle < e^{-c/6}15/\sqrt{2}$ during inflation.
Hence, we infer that the first term of Eq.~(\ref{correctionstonn}) exerts the dominant impact on inflationary dynamics.
With the inclusion of this term in the K\"ahler potential, the scalar potential Eq.~(\ref{scalarpot}) is altered to ($\langle K\rangle=c$ is assumed)
\begin{align} 
V=& \ \exp\left[\sum_{i=1}^3 e^{c/3} \ \lambda_i \tilde{N}_i^{'\dagger}\tilde{N}'_i\right]
\nonumber\\
&\times\sum_{i=1}^3\frac{1}{1+\lambda_i}\left\{e^{2c/3}\left\vert\frac{\partial W}{\partial \tilde{N}'_i}\right\vert^2
+e^c\,\lambda_i\left(\frac{\partial W}{\partial \tilde{N}'_i}\tilde{N}'_i W^\dagger+{\rm H.c.}\right)+e^{4c/3}\,\lambda_i^2\vert W\vert^2\tilde{N}_i^{'\dagger}\tilde{N}'_i\right\}, 
\label{altscalarpot}\\
&{\rm with} \ \lambda_i=\frac{\log\Lambda^2}{32\pi^2}[hh^\dagger]_i \ ,
\nonumber
\end{align}
 where $\tilde{N}'_i$ span the flavor basis where $(hh^\dagger)_{ij}$ is diagonal,
 and $[hh^\dagger]_i$ denotes the eigenvalue of $(hh^\dagger)_{ij}$ belonging to $\tilde{N}'_i$.
In the current model, we may approximate $\tilde{N}'_i\simeq \tilde{N}_i$ and $[hh^\dagger]_i\simeq(hh^\dagger)_{ii}$.

During inflation, the transplanckian VEV of $\phi_3$ induces sizable mass terms for $\phi_1,\phi_2,\phi_3$ through $\lambda_i$ couplings
 in Eq.~(\ref{altscalarpot}).
For $\phi_3$, the induced mass term, $m_{3,{\rm induced}}$, is estimated to be
\begin{align} 
m_{3,{\rm induced}}^2&\simeq \frac{1}{2}M_3^2\phi_3^2
\left\{
(29\lambda_3+79\lambda_3^2\phi_3^2+28\lambda_3^3\phi_3^4+2\lambda_3^4\phi_3^6)
\exp\left[\frac{\lambda_3}{2}\phi_3^2\right]+2\frac{\exp\left[\frac{\lambda_3}{2}\phi_3^2\right]-1}{\phi_3^2}\right\}
\nonumber \\
&=3H^2\left\{
(29\lambda_3+79\lambda_3^2\phi_3^2+28\lambda_3^3\phi_3^4+2\lambda_3^4\phi_3^6)
\exp\left[\frac{\lambda_3}{2}\phi_3^2\right]+2\frac{\exp\left[\frac{\lambda_3}{2}\phi_3^2\right]-1}{\phi_3^2}\right\},
\end{align}
 where $H$ denotes the Hubble rate and we have used the fact that the vacuum energy almost equals $M_3^2\phi_3^2/2$.
Since we have $(hh^\dagger)_{33} \lesssim 2(\vert\Delta m^2_{23}\vert+\Delta m^2_{12})M_3/v^2 < 0.04$ in the viable parameter region of our scenario, and since $\phi_3<15$ holds during inflation,
 the induced mass is bounded from above as 
\begin{align} 
m_{3,{\rm induced}}<0.17H \ \ \ {\rm for} \ \log\Lambda^2=2.
\end{align}
For $\phi_1$ and $\phi_2$, their induced masses are below $m_{3,{\rm induced}}$ (note $\lambda_1<\lambda_2<\lambda_3$).
We have thus confirmed that for a reasonable value of the cutoff scale, such as $\log\Lambda^2=2$,
 the induced mass terms for $\phi_1,\phi_2,\phi_3$ during inflation are below $0.17H$ and thus do not affect their slow-roll.
\\

\section{Summary}
We have investigated a scenario where the supersymmetric partners of two right-handed neutrinos (sneutrinos) work as mixed inflaton and curvaton.
We have estimated the widths of the sneutrinos (=inflaton and curvaton) from the measured active neutrino mass differences,
 under natural assumptions on the lightest neutrino mass and the rotation matrix $R_{3\times3}$ in Casas-Ibarra parametrization.
Through a numerical calculation, we have demonstrated that the tensor-to-scalar ratio $r$ can be reduced so that the scenario
 is made consistent with the BICEP2/Keck Array and Planck data.
We have further derived a specific prediction on local non-Gaussianities $f_{\rm NL}$ and $\tau_{\rm NL}$
 in terms of the second heaviest sneutrino mass $m_\sigma$ and its VEV when the pivot scale exited the horizon $\sigma_*$,
 based on the sneutrino widths estimated above.
This prediction is compared to the Planck bound on $f_{\rm NL}$ to constrain $m_\sigma$ and $\sigma_*$.
Also, it has been revealed that future measurements of local non-Gaussianities with an accuracy of $\Delta f_{\rm NL}\lesssim1$
 possibly give strong restrictions on $m_\sigma$ and $\sigma_*$.
Interestingly, the allowed regions in the $\sigma_*$--$m_\sigma$ plane and the prediction on non-Gaussianities 
 are different for the cases with the normal and inverted hierarchy of the active neutrino mass.
In contrast to the strong correlation with neutrino physics, it is not easy to relate our results with collider searches of supersymmetric particles, because the mechanisms for supersymmetry breaking and its mediation to the visible sector are not relevant in our scenario and hence soft supersymmetry breaking masses may take any values.
\\

\section*{Acknowledgement}
T.T. would like to thank  Daisuke Yamauchi for useful correspondence on the future constraint on $\tau_{\rm NL}$.
This work is partially supported by Scientific Grants by the Ministry of Education, Culture, Sports, Science and Technology of Japan (Nos. 24540272, 26247038, 15H01037, 16H00871, and 16H02189) (NH),  JSPS KAKENHI Grant Number 15K05084  (TT), 17H01131 (TT), and MEXT KAKENHI Grant Number 15H05888 (TT).
\\

\end{document}